\documentclass[twocol]{ametsocV6.1}

\usepackage{accents}
\usepackage{tabularray}

\defcitealias{coffer2019srh500}{C19}
\defcitealias{coniglio2020insights}{CP20}
\defcitealias{coniglio2022mesoanalysis}{CJ22}
\defcitealias{coniglio2024sampling}{CT24}

\title{Supercell environments using GridRad-Severe and the HRRR: Addressing discrepancies between prior tornado datasets}

\authors{Brice E. Coffer\aff{a}\correspondingauthor{Brice E. Coffer, becoffer@ncsu.edu}\thanks{Current affiliation: European Severe Storms Laboratory, Wiener Neustadt, Austria}, Matthew D. Parker\aff{a}, 
Michael C. Coniglio\aff{b}\aff{c}, Cameron R. Homeyer\aff{c}}
\affiliation{\aff{a}{Department of Marine, Earth, and Atmospheric Sciences, North Carolina State University, Raleigh, NC}\\
\aff{b}{NOAA/National Severe Storms Laboratory, Norman, OK}\\
\aff{c}{School of Meteorology, University of Oklahoma, Norman, Oklahoma}\\
}

\abstract{Storm-relative helicity (SRH) is an important ingredient in supercell development, as well as mesocyclone intensity, and is linked to tornadogenesis and tornado potential. Derived from the storm-relative wind profile, SRH is composed of both the vertical wind shear and storm-relative flow. Recent studies have come to conflicting findings regarding whether shallower or deeper layers of SRH have more skill in tornado forecasting. Possible causes of this discrepancy include the use of observed versus model-based proximity soundings, as well as whether the storm-relative wind profile is determined via observed versus estimated storm motions. This study uses a new dataset of objectively identified supercells, with observed storm motions, paired with high-resolution model analyses to address the discrepancies among prior studies. Unlike in previous model-based tornado environmental datasets, the present approach reveals substantive differences in storm-relative flow, vertical wind shear, and SRH within the low-to-mid-levels between nontornadic and tornadic supercells. Using observed storm motions for storm-relative variables further magnifies differences in the low-to-mid-level storm-relative winds between nontornadic and tornadic supercells, ultimately leading to deeper layers of SRH having more forecast skill than near-ground SRH. Thus, the combination of a higher-resolution model analysis, which better represents the near-storm environment, with observed storm motions appears to explain why many past tornado climatologies using model-based environmental analyses have failed to find significant differences in the storm-relative wind profile. These results help bridge the gap between previous studies that employed coarser model-based analyses and those that aggregated observed soundings from field projects.}

\begin{document}

\maketitle

%
%

%

\section{Introduction}

Our understanding of environmental characteristics that favor severe convective weather, including why some supercell thunderstorms produce tornadoes and others do not, has improved considerably in recent decades \citep{markowski2009tornadogenesis,davies2015review,fischer2024progress} and continues to undergo refinements based on new datasets and field observations \citep[e.g.,][hereafter \citetalias{coffer2019srh500} and \citetalias{coniglio2020insights}, respectively]{coffer2019srh500,coniglio2020insights}. When combined with improvements in numerical weather prediction, especially in the mesoscale regime \citep[e.g.,][]{james2022hrrr}, as well as the widespread adoption of convection allowing models and ensembles \citep[e.g.,][]{roberts2022href}, regional forecasts for severe weather show a continuous rate of improvement \citep{hitchens2014evaluation,herman2018probabilistic}. 

The climatological studies of environmental tornado proxies that have been used to improve these severe weather forecasts generally fall into two categories: those that use observed soundings \citep[e.g.,][]{fawbush1952mean,beebe1958tornado,maddox1976evaluation,johns1992severe,brooks1994environments,rasmussen1998baseline,rasmussen2003refined,craven2004baseline,parker2014composite,wade2018comparison} and those that use model-based analyses \citep[e.g.,][]{thompson1998eta,thompson2003close,thompson2007effective,thompson2012convective,markowski2003characteristics,togstad2011conditional,nowotarski2013classifying,anderson2016investigation} to represent the near-storm environment. Each category has its pros and cons. Observed soundings have the benefit of being direct measurements of the current state of the atmosphere, but historically have had to compromise on temporal and spatial proximity to populate a large sample size \citep{potvin2010assessing}. The presence of strong spatial gradients, combined with rapid temporal evolution and storm-scale feedbacks \citep[e.g.,][]{markowski1998variability,nowotarski2014properties,parker2014composite,davenport2015impact,wade2018comparison}, make it challenging for the operational upper-air network to adequately represent supercell near-storm environments. Using multiple decades of observed soundings from field projects, \citetalias{coniglio2020insights} addressed many of these limitations. However, field project operations are most often undertaken during the spring months, in the central region of the United States, and tend to focus on robust, high-end supercells \citepalias{coniglio2020insights}; thus, they may not represent the full supercell spectrum \citep[hereafter \citetalias{coniglio2024sampling}]{coniglio2024sampling}. On the other hand, the primary benefit of model-based analyses is their superior temporal and spatial coverage compared to field projects and the operational upper-air sounding network. With a greater number of cases available for analysis comes the drawback of relying on the model's representation of the environment, which is known to have biases \citep[hereafter \citetalias{coniglio2022mesoanalysis}]{coniglio2012verification,coniglio2022mesoanalysis}, especially in the planetary boundary layer. 

Recent research, using both observed \citep{parker2014composite,coffer2017simulated,coffer2018tipping} and model-based analyses \citep[\citetalias{coffer2019srh500},][]{coffer2020era5,nixon2022distinguishing,goldacker2021updraft,goldacker2023srh}, has focused on how characteristics of the near-ground wind profile influence the organization and intensity of the low-level\footnote{We define low-levels as approximately 1000 m above ground level (AGL), typically at or below the level of free convection. This is in contrast to near-ground ($<$500 m AGL) and mid-levels ($\sim$3--6 km AGL). Accordingly, low-to-mid-levels is defined as 1--3 km AGL.} mesocyclone in supercells. Near-ground environmental streamwise horizontal vorticity, and its storm-relative flux into an updraft (storm-relative helicity; SRH), has been found to be skillful at distinguishing between nontornadic and significantly tornadic supercells that produce EF2+ tornadoes \citep[e.g.,][\citetalias{coffer2019srh500}, \citetalias{coniglio2024sampling}]{parker2014composite}. In numerical cloud modeling focused on the lower tropospheric wind profile, highly streamwise near-ground wind profiles yield organized, steady low-level mesocyclones, increasing the likelihood of tornadogenesis \citep[e.g.,][]{markowski2014influence,coffer2017volatility,peters2023disentangling}. Air ingested into the low-level mesocyclone almost exclusively comes from near the ground \citep{coffer2023LLM}, crucially providing the dynamic lifting needed for tornadogenesis below the level of free convection where buoyancy cannot contribute (distinguishing supercells from ordinary thunderstorms).

While the influence of the near-ground wind profile on the low-level mesocyclone is clear, the specific claim that the near-ground SRH (e.g., 0--500 m SRH, SRH500) is statistically more skillful than SRH computed over deeper layers \citep[e.g., 0--3 km SRH, SRH3; or SRH within the effective inflow layer, ESRH;][]{thompson2007effective} has since been questioned. Using observed, field project soundings, \citetalias{coniglio2020insights} found substantial differences in the low- to mid-level wind profile from 1--3 km AGL, both in the vertical wind shear and storm-relative winds. The combination of these components yields larger SRH when integrated over deeper layers and more forecast skill distinguishing tornadic/significantly tornadic supercells from their nontornadic counterparts. These differences have not been noted in any model-based tornado environment studies to date \citep[][C19]{markowski2003characteristics,thompson2003close,thompson2012convective}, which have all been based on analyses from of the Rapid Update Cycle \citep[RUC;][]{benjamin2004ruc} and the Rapid Refresh \citep[RAP; after 1200 UTC 1 May 2012;][]{benjamin2016rap}. 
A physical explanation for the role of the low-to-mid-level wind profile is still yet to be clarified in the literature, but could involve the distribution of hydrometeors \citep{brooks1994role,parker2017backing,warren2017impact,gray2021midlevel}, the size of the storm and related updraft dilution \citep{peters2020updraft,mulholland2024vws}, the intensity/longevity of the mid-level mesocyclone \citep[and the vertical alignment with the low-level mesocyclone;][]{homeyer2020distinguishing} modulating the vertical velocity at lower altitudes, or the structure of horizontal rotors and the resultant pressure field in vicinity of the updraft \citep{muehr2024influence}, amongst others. 

\begin{figure*}[t]
\centerline{\includegraphics[width=40pc]
{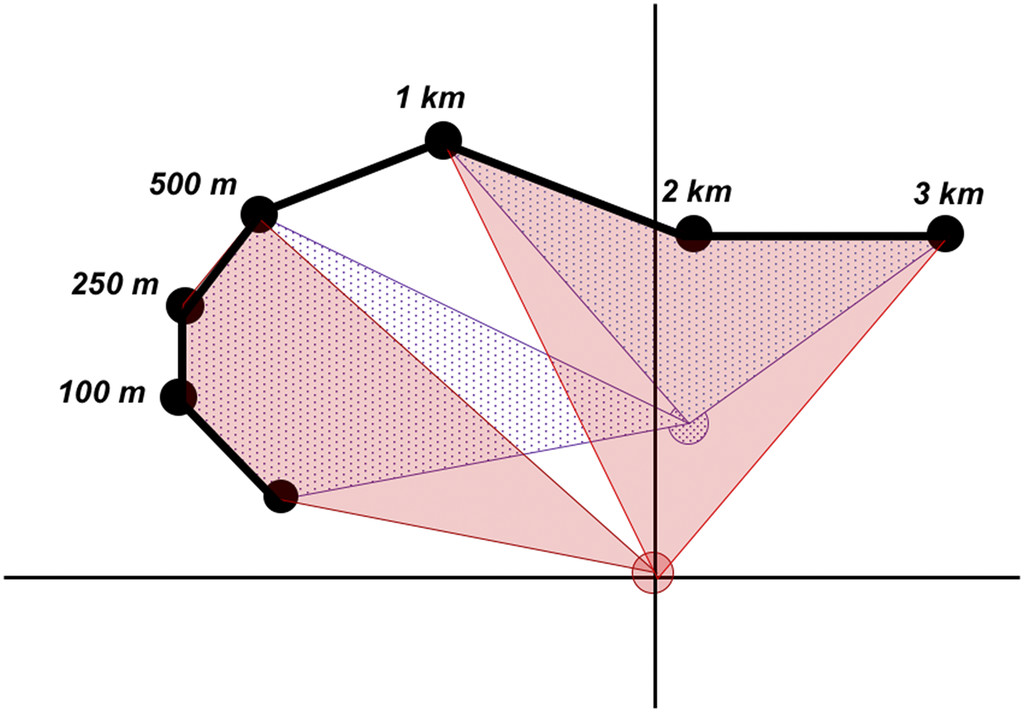}}
\caption{Illustration of how SRH computed over 0–500 m and 1–3 km changes with a more rightward-deviant storm motion. The red-shaded (purple-hatched) areas depict SRH computed over 0--500 m and 1--3 km resulting from the observed (Bunkers) storm motion. Reproduced from \citet{coniglio2020insights} with permission and \copyright American Meteorological Society.}
\label{fig:SRHexample}
\end{figure*}

Another key difference between the \citetalias{coffer2019srh500} and \citetalias{coniglio2020insights} studies is the use of observed storm motions in \citetalias{coniglio2020insights} compared to storm motion estimates using the Bunkers method \citep{bunkers2000predicting} in \citetalias{coffer2019srh500} and most other model-based climatologies. The Bunkers method is an empirical estimate of storm motion solely based on the wind-profile and has no knowledge of storm-scale conditions, such as neighboring convection or the convection initiation boundary orientation. Observed tornadic storms in \citetalias{coniglio2020insights} deviated more to the right of the mean wind than empirical estimates would suggest, especially in larger SRH environments \citep[consistent with the results of][] {bunkers2018observations}. Most notably, tornadic storms in the \citetalias{coniglio2020insights} dataset tended to temporarily deviate rightward in the 25 minute period prior to tornadogenesis. The deviation in observed storm motion affects whether shallower or deeper layers of SRH yield more forecast skill, as the magnitude of the storm-relative winds tends to be enhanced most in the layer above 500 m AGL, while staying about the same in the 0--500 m layer (illustrated conceptually in Fig.~\ref{fig:SRHexample})\footnote{This conceptual illustration is most relevant to hodographs typical of springtime, Great Plains severe weather events, which comprise the majority of the \citetalias{coniglio2020insights} dataset. For other regimes, such as cool-season, southeastern United States severe weather events, where the low-level winds are often stronger and more veered (i.e., the hodograph is entirely in the upper-right quadrant), then a rightward deviation in storm motion would also enhance the near-ground SRH}. Whether this short time scale deviant motion is a cause or effect of a storm's propensity to produce a tornado is yet unknown. 

In summary, discrepancies between tornado climatologies constructed with observed soundings versus those from model analyses have led to conflicting interpretations of whether shallower layers of SRH are more or less skillful at forecasting significant tornadoes than deeper layers. Questions remain over whether observed datasets from field projects capture the full spectrum of supercells due to being more restricted seasonally and diurnally, as well as being biased towards particularly strong supercells. It is also still unclear how important observed (vs. estimated) storm motions are in terms of their impact on the storm relative winds. \citetalias{coniglio2024sampling} found that restricting model analyses to a spatiotemporal region that emulates field projects did not explain the disagreement; however, they noted that the tendency for field projects to target particularly strong storms, and the use of actual storm motions versus estimates of storm motions, both contribute to the conflicting results.

In this companion study to \citetalias{coniglio2024sampling}, we further explore the conflicting recent results between model-based and observed supercell proximity sounding datasets using an independent storm dataset, including observed storm motions, derived from GridRad-Severe \citep[hereafter abbreviated GR-S;][]{gridradsevereCISL} and model analyses from the High Resolution Rapid Refresh \citep[HRRR;][]{dowell2022hrrr}, a convection-allowing model tailored to representing finer scale weather phenomena. The unique aspects of the present study are using the HRRR for near-storm environmental analyses---potentially more accurately portraying finer-scale heteorgeneties and storm-induced modifications in the near-storm environment---as well as GR-S's ability to generate a large population of objectively tracked supercells of varying intensity, across geographically diverse regions. These aspects differentiate the present work from prior model-based climatologies and those from observational field projects. Specifically, we employ those new data sources to ask the following research questions:

\begin{enumerate}
    \item Is the representation of the near-storm environment in HRRR analyses more similar to observations from field projects or to tornado climatologies with traditionally used, coarser model-based analyses?
   \item If the HRRR analysis is indeed more like previous observational datasets, does the representation of the low-to-mid-level wind profile clarify prior discrepancies in whether shallower or deeper layers of SRH have more forecast skill?
    \item How much do observed storm motions (compared to Bunkers storm motion estimates) alter the forecast skill of storm-relative variables within a large, climatologically representative storm sample?
\end{enumerate}

This work and \citetalias{coniglio2024sampling} share some similarities in the stated research goals, albeit using distinct methods to tackle the questions at hand. The primary goal across both works is to bridge the gap between model-based and observed tornado climatologies by either corroborating or rejecting environmental traits previously identified as having more forecast skill. 

The rest of the paper is organized as follows. Details describing the respective supercell climatologies and model analyses are outlined in Section~\ref{methods}. Direct comparisons between analyses derived from a traditionally used, coarser model and a more modern, higher resolution model for the same subset of events are in Section~\ref{HRRRvsSFCOA}. Comparisons between a widely used supercell climatology compiled by the Storm Prediction Center (SPC) and a new independent, objectively identified dataset of supercells from GR-S are in Section~\ref{GRSvsSPC}, with the impact of observed storm motions within this new dataset discussed in Section~\ref{GRSvsSPC-SM}. A summary of the main conclusions and a discussion of outstanding research questions are in Section~\ref{Conclusions}. 

\section{Methods}\label{methods}

\subsection{GR-S HRRR-Inflow dataset}

An independent dataset of supercell thunderstorms was compiled using GR-S. GR-S matches severe weather reports and storm tracks with a publicly available gridded radar dataset, GridRad \citep{GridRaddataset}. The technical details behind the creation of GridRad and GR-S can be found in \citet{GridRaddataset} and \citet{murphy2023gridrad}, respectively. Some of the most relevant details from those studies to the current work are repeated below. 

GR-S event days are defined as high-end severe weather days where the number of tornado, hail, or wind reports exceed 8, 45, or 120, respectively. These thresholds were chosen to satisfy the need for a balanced dataset across all hazard types while managing the overall size of the generated data. Because of this, however, the dataset of supercell storms are likely slightly skewed towards days with more numerous storm events, which might bias environmental characteristics in unknowable ways. Once event days are identified, storm tracks are identified using a tracking method from \citet{homeyer2017cirrus} with modifications from \citet{lagerquist2020deep}. The GridRad tracking algorithm specifically identifies 30-dBZ echo top altitude maxima higher than at least 4 km above mean sea level (AMSL). This altitude threshold could have implications for the identification of cool season convection, such as in high-shear, low-CAPE regimes, where the tropopause height is lower and convection can be quite shallow \citep[$<$ 4 km AMSL;][]{davis2014radar}. Finally, from this tracking algorithm, each GR-S event records a 30-min moving average storm motion based on the latitude and longitude of the echo top maxima throughout the storm's lifecycle. 

\begin{figure*}[t]
\centerline{\includegraphics[width=30pc]{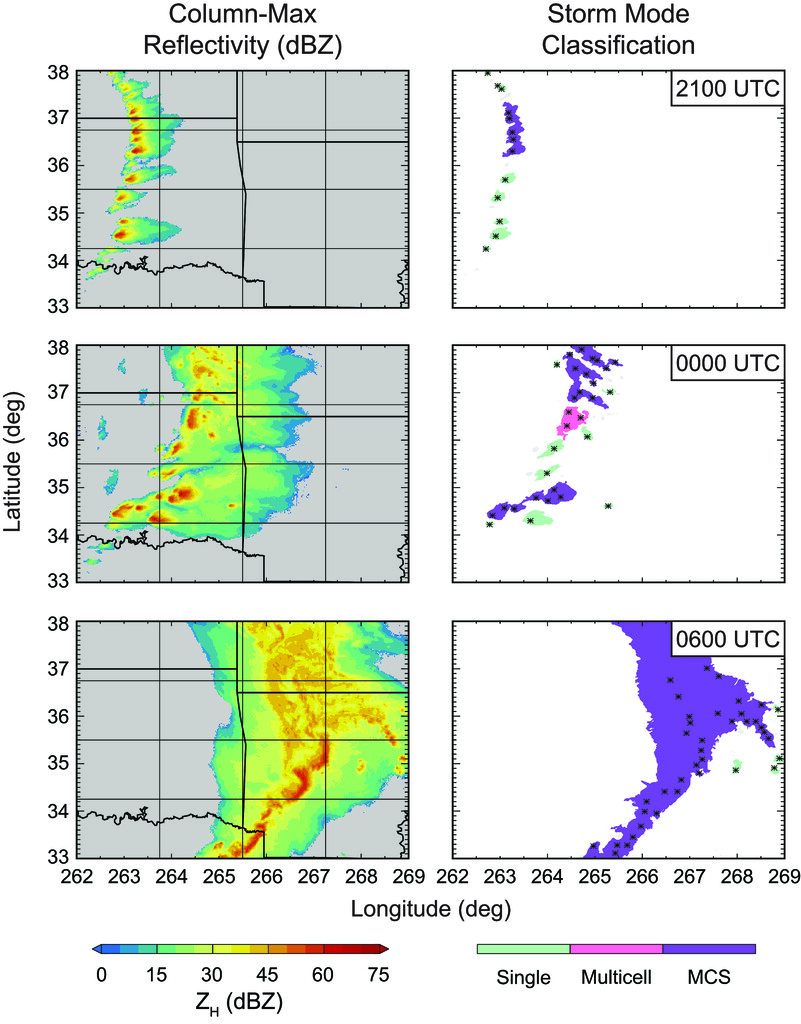}}
\caption{Select images of (left) column-max reflectivity and (right) storm mode classification from a limited spatial domain within the 14 and 15 Apr 2011 GR-S case. For storm mode classification images, the interior of identified 30-dBZ contours are shaded in green, pink, or purple to denote single-cell, multicell, or MCS classification, respectively. Tracked storms within such contours are indicated by black asterisks. Identified 30-dBZ contours that do not encompass any storm tracks are shaded in light gray. Reproduced from \citet{murphy2023gridrad} with permission and \copyright American Meteorological Society.}
\label{fig:GR-S-example}
\end{figure*}

GR-S further delineates between storm modes based on thresholds in the 30-dBZ contours of column-maximum reflectivity. Storm tracks are uniformly binned into three storm modes: single cell, multicell, or mesoscale convective system (MCS), based on the size of the 30-dBZ contour and/or the number of tracked echo tops within this contour. On the whole, these criteria effectively capture the typical evolution of severe convection (Fig.~\ref{fig:GR-S-example}), but they likely lack the subtlety and detail of the convective mode classification performed manually by \citet{smith2012modes} and used by \citetalias{coffer2019srh500}. GR-S has a tendency to merge nearby cells that would likely be classified as single-cell or multicells using base-scan reflectivity, but whose column-max 30-dBZ reflectivity contours merge (e.g., the storms in northern Oklahoma at 2100 UTC in Figure~\ref{fig:GR-S-example}). As a result, the distribution of severe weather events tends to skew towards MCS storm modes [as noted by \citet{murphy2023gridrad} and discussed further below]. 

To identify supercells specifically, several radar based ``mesocyclonic updraft'' criteria were applied to each storm track beyond what is provided by GR-S. These five criteria are from \citet{homeyer2020distinguishing}, based on prior work by \citet{sandmael2017evaluation}: a supercell is defined to have 1) maximum 4--7 km AMSL azimuthal shear exceeding 4 x 10$^{-3}$ s$^{-1}$ for at least 40 min in total over the entire storm lifetime; 2) maximum 4--7 km AMSL azimuthal shear meeting or exceeding 5 x 10$^{-3}$ s$^{-1}$; 3) maximum column-max azimuthal shear meeting or exceeding 7 x 10$^{-3}$ s$^{-1}$; 4) maximum radial divergence at any altitude meeting or exceeding 1 x 10$^{-2}$ s$^{-1}$; 5) maximum velocity spectrum width at any altitude meeting or exceeding 13 m s$^{-1}$. \citet{homeyer2020distinguishing} also used a sixth criterion for maximum 40-dBZ echo top altitude of 11 km AMSL. 
In the present study, the 40-dBZ height threshold was lowered from 11 km to 6 km for nontornadic storms and eliminated all together for tornadic storms \citep[following][]{murphy2023gridrad} to identify more cool season convection (admittedly with partial success, as shown later in Section~\ref{GRSvsSPC}a). The differing criteria for nontornadic and tornadic storms were a compromise to limit the nontornadic cases to a reasonable number, while maximizing the number of tornadic cases.

Storm event tracks from GR-S over the period spanning 12 UTC on 23 August 2016 to 31 December 2023, terminating at 2359 UTC, were considered. This time frame corresponds to the overlap between the implementation of HRRRv2 or later\footnote{Environmental fields from HRRRv1 were subjectively deemed to be of lower quality, especially near assimilated convection, and not suitable for this purpose.} and the current availability of GR-S. The storm mode and mesocyclone filters were applied to storm event tracks to identify relatively isolated convection with mesocyclonic rotation. GR-S continuously updates the storm mode classification along each storm track based on the area of column-max reflectivity. Nontornadic storms were strictly required to be single or multicell (not MCS). To maximize total tornadic sample size and attempt to catch events on the edge of the single/multicell and MCS classes, tornadic storms were required to have at least one instance of being classified as a single or multicell within a window 15 minutes prior to the formation of the maximum intensity tornado until the demise of that tornado. While exploring the GR-S dataset, we found what we considered isolated single- or multi-cell storms were occasionally labeled as MCSs by GR-S (a result of the necessary ``one size fits all'' thresholds applied in GR-S; Fig.~\ref{fig:GR-S-example}). For all tornadic MCS cases meeting the five mesocyclonic updrafts criteria ($\sim$1200 events), storm mode was manually revisited in a process similar to \citet{smith2012modes}, focusing on discrete, cell-in-cluster, and cell-in-line right-moving supercells. Of these tornadic MCS cases, our manual classification determined roughly 5\% were isolated supercells and a further 15\% were cell-in-line or cell-in-cluster supercells. These cases were included in the final analysis. Most of the more egregious `misses' from GR-S's objective storm mode classification were supercells ahead of an MCS, high-precipitation supercells, lines of convection developing along a front/dryline, messy convective modes (but not organized into a MCS), or clusters of storms at long ranges from the nearest radar. The low number of cases that were manually considered `misclassified' speaks to the robustness of the thresholds applied in \citet{homeyer2020distinguishing} and \citet{murphy2023gridrad}. In general, because the sample primarily constitutes the single and multicell storm mode labels, we are confident that the supercell storms are mostly isolated in nature (similar to cases in the SPC dataset analyzed in \citetalias{coffer2019srh500} and cases often targeted by field projects, as in \citetalias{coniglio2020insights}). 


Environmental fields from the HRRR on the native (sigma) model levels were pulled from the Amazon Web Server HRRR archive (\url{https://registry.opendata.aws/noaa-hrrr-pds/}) using the Herbie package \citep{blaylock2017hrrr,blaylock2022herbie}. The event time for nontornadic storms was at the time of maximum low-level azimuthal shear (0--2 km AMSL) from each GR-S track (if multiple instances of this value occurred, only the first time was considered). The event time for tornadic storms was defined as the start time of the most intense tornado during the storm track. If a single storm track contained multiple instances of equal maximum tornado intensity (i.e., multiple EF1 tornadoes but no EF2+), then the first maximum intensity tornado track was used. Only one tornado per storm track was considered to ensure that prolific tornadic storms are not being repeatedly sampled. Environmental fields were extracted from the 00-h HRRR analysis immediately prior to the time of storm event (i.e., for an event time at 2347 UTC, the HRRR analysis at 2300 UTC is used). Similar to \citet{thompson2012convective}, the prior analysis time is preferred over the closest analysis time in order to help ensure the environmental profile is less likely to be within or behind the storm. 

Prior studies using model-based analyses to construct large supercell environment datasets have tended to use regional or global models with relatively coarse horizontal grid spacing (10–40 km). Point environmental profiles from mesoscale models with convection allowing grid spacings might be less likely to be representative of the mean base-state experienced by a storm because of the model's ability to resolve finer-scale processes (this possibility is explored over the common parameter space of supercell environments in Section~\ref{HRRRvsSFCOA}). 
On the other hand, it is also reasonable to assume convection-allowing models, which directly assimilate storm-scale features, may have value in representing environmental heterogeneities and storm-scale modifications not possible in coarser, regional models \citep{kerr2019diagnosing,potvin2019systematic,potvin2020assessing,laser2022doppler}. Striking the right balance between these details is an important consideration because a representative inflow environment is paramount to any environmental tornado proxy study \citep{potvin2010assessing}. As forecasters increasingly rely on higher resolution analyses \citep{wade2023regional}, it is also important to evaluate the representation of tornado environmental proxies using these newer analyses. Preliminary attempts by \citetalias{coffer2019srh500} to extract environmental proxies from the HRRR were hindered by substantial difficulty in defining a representative environment in an automated fashion over a large number of cases. 

\begin{figure*}[t]
\centerline{\includegraphics[width=40pc]{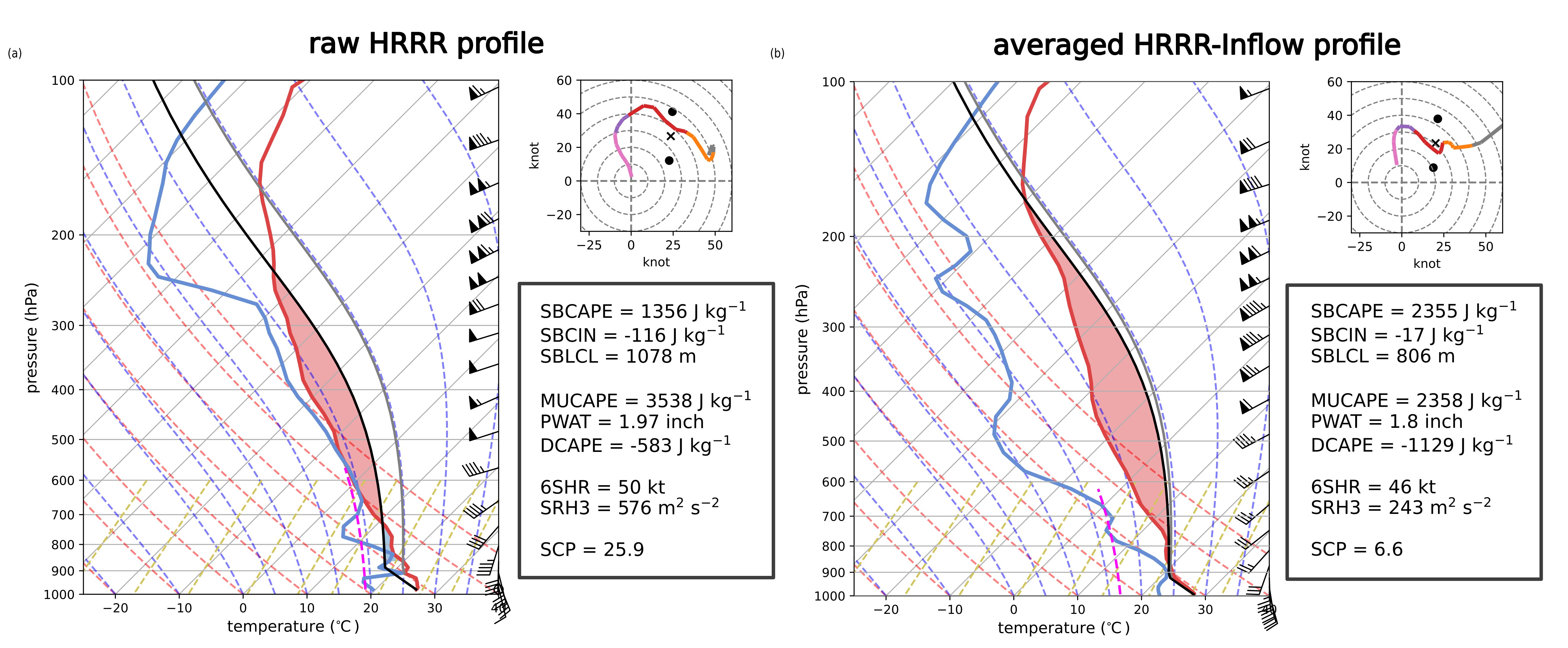}}
\caption{Skew$T$ log-$p$ diagrams, with accompanying hodographs, of example near-storm environmental profiles of the (left) raw HRRR point profile versus (right) the HRRR-Inflow averaged inflow profile for a significantly tornadic supercell near Dallas, TX on 21 October 2019 at 02Z. Plotted on the Skew$T$ log-$p$ diagrams is temperature (red), dewpoint temperature (blue), surface-based parcel path (black) with areas of positive and negative buoyancy shaded in red and blue, respectively, and the most-unstable parcel path (light gray). On the hodograph diagrams, the 0--500 m, 500 m--1 km, 1--3 km, 3--6 km, and 6--9 km layers are highlighted in pink, purple, red, orange, and gray, respectively, while the 0--6 km mean wind is represented by a cross and the left and right Bunkers storm motion estimate are represented by filled circles. All listed values of CAPE were calculated using the parcel’s virtual temperature.}
\label{fig:soudings_raw_hrrrinflow}
\end{figure*}

\begin{figure*}[t]
\centerline{\includegraphics[width=40pc]{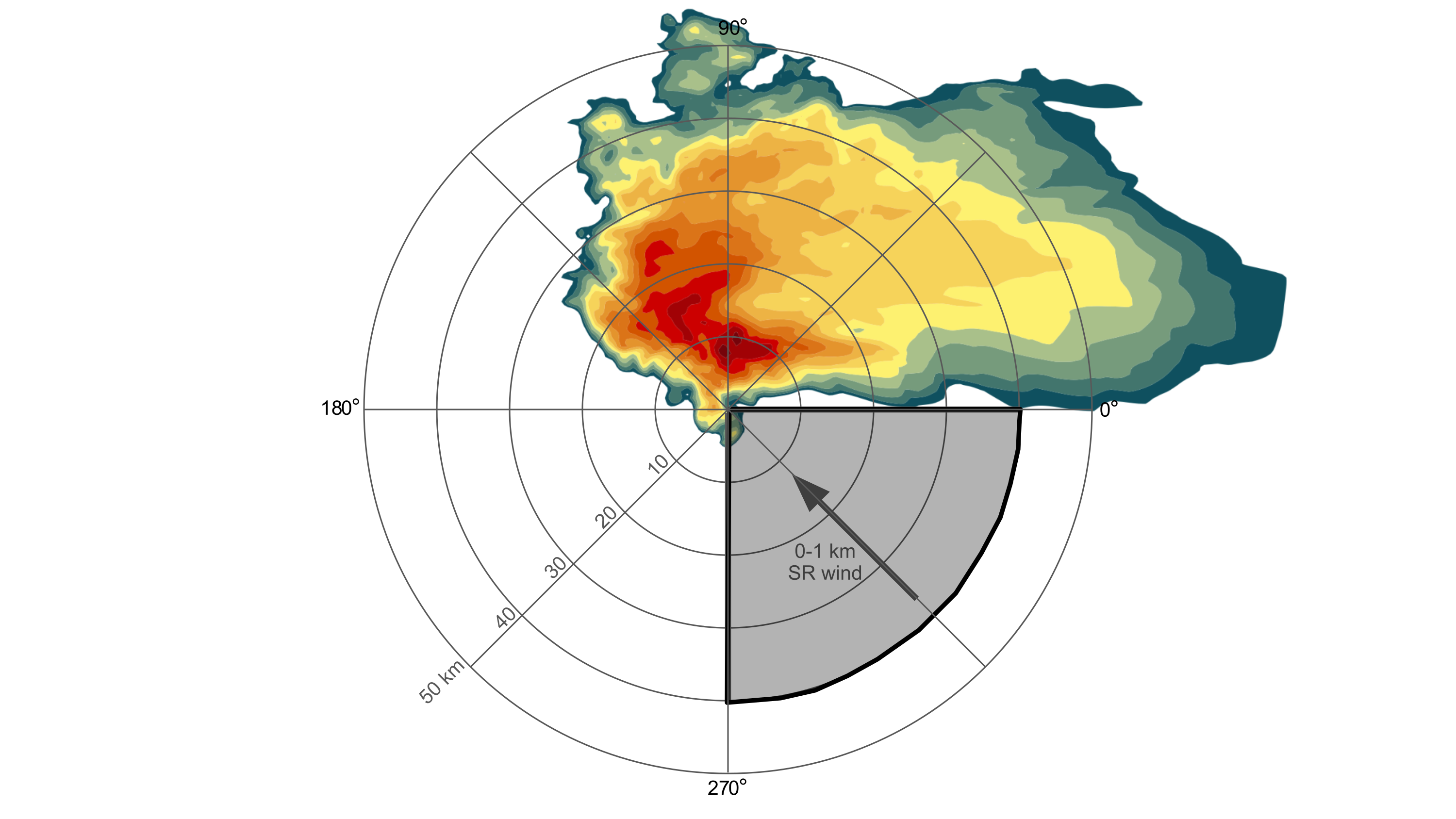}}
\caption{Conceptual diagram of the area (shaded; gray) where environmental state variables are averaged within a 40 km, 90$^{\circ}$ ``inflow'' slice for the HRRR-Inflow averaged inflow profile in Figure~\ref{fig:soudings_raw_hrrrinflow}b. The 0--1 km AGL storm-relative wind vector is directed towards the SE at 315$^{\circ}$ and environmental state variables are averaged 45$^{\circ}$ either side of this vector. For reference, the raw HRRR point profile from Figure~\ref{fig:soudings_raw_hrrrinflow}a would be extracted at the origin and roughly 140 grid points in the HRRR comprise the inflow average.}
\label{fig:hrrrinflow_concept}
\end{figure*}

To this end, the present study employs a spatial averaging technique inspired by the `calibrated inflow' machine learning techniques \citep{jahn2020inflow} used during the NOAA Hazardous Weather Testbed Spring Forecasting Experiment \citep{clark2023sfe}\footnote{Preliminary results from the Spring Forecasting Experiment suggest calibrated tornado forecasts using environmental data from within the ``inflow'' of the near-storm environment result in slightly better forecast skill (\url{https://hwt.nssl.noaa.gov/sfe/2022/docs/HWT_SFE_2022_Prelim_Findings_FINAL.pdf}).}. For each GR-S storm event location, a vertical profile of preliminary atmospheric state variables ($\theta$, $q_v$, $p$, $u$, $v$, $w$, $z_{AGL}$) was taken from the nearest neighbor HRRR grid-point along the GR-S event track. Oftentimes this raw vertical profile contained unrepresentative features 
(Fig.~\ref{fig:soudings_raw_hrrrinflow}a), such as low-level cooling and drying, saturated moist adiabatic layers, and/or large, looping hodographs. To alleviate this, from the grid-point wind profile, the 0-1 km storm-relative wind direction based on Bunkers storm motion estimate\footnote{Using the Bunkers storm motion potentially introduces a source of error compared to using the actual storm motion provided by GR-S; however, for the comparisons in Section~\ref{HRRRvsSFCOA}, actual storm motions were not available. Thus, for consistency across the entire study, Bunkers storm motion was preferred. If future studies use a similar inflow-averaging technique, and actual storm motion are available, it could be valuable to compare the effects of storm motion on the averaging process.} is calculated and an inflow region is defined to extend 45$^{\circ}$ in either direction of the low-level storm-relative wind vector within a 40 km radius (Fig.~\ref{fig:hrrrinflow_concept}). This 90$^{\circ}$ slice is representative of typical inflow trajectories into the low-level mesocyclone across a range of supercells in \citet{coffer2023LLM}. The same atmospheric state variables are then averaged within the 90$^{\circ}$ slice to generate a mean inflow sounding\footnote{While fields at very close distances to the storm event point (i.e., $<$ 10 km) are still part of the averaging process, this area represents a small portion of the overall ``inflow'' average ($\sim$ 7\%). Fields beyond 40 km were also considered but were subjectively deemed increasingly detrimental to the averaged sounding, mainly due to increasing convection inhibition.} (Fig.~\ref{fig:soudings_raw_hrrrinflow}b). Hereafter, this averaged inflow sounding is termed HRRR-Inflow.

This procedure automates the generation of environmental proxies from the HRRR, while minimizing some of the undesired effects of convection-allowing grid-spacing. A possible drawback of this approach is that the initial, raw grid-point's 0-1 km storm-relative wind could be unrepresentative of the actual inflow region (especially if the direction is inaccurate), contaminating the average inflow sounding. All large model-based tornado climatologies have the caveat that at least some proxy soundings are not representative of a storm's actual environmental fields. As in previous studies, these issues should be alleviated by using a large number of cases and by using the prior analysis time in the HRRR relative to the time of the event. 

To further remove environments unlikely to be representative of surface-based supercellular convection, soundings with zero most unstable (MU) convective available potential energy (CAPE), an effective inflow base (EIB) above the surface, or an effective bulk wind difference (EBWD) less than 12.5 m s$^{-1}$ (the lower limit of EBWD in the formulation of the Significant Tornado Parameter; STP) were removed from consideration (as in \citetalias{coniglio2024sampling}). In total these filters removed 4093 cases ($~$40\%), most of which were nontornadic events ($>$95\%).

In summary, using GR-S storm tracks from late 2016 to 2023 and filtering for relatively isolated storms with mesocyclonic updrafts, an independent supercell dataset with environmental data from a convection-allowing model was generated to compare to previous model-based studies.  This filtering, combined with the availability of archived HRRR data, resulted in a total of 6397 nontornadic supercells (both non-severe and severe), 764 weakly tornadic supercells (EF0-1), and 255 significantly tornadic supercells (EF2+). Additionally, for these cases, both Bunkers estimated storm motion and actual storm motion vectors are available to compute storm-relative quantities. While the total case counts are less than the broader dataset in \citetalias{coffer2019srh500}, the sample size of weakly tornadic and significantly tornadic events is substantially more than in \citetalias{coniglio2020insights} and is similar to prior studies, such as \citet{markowski2003characteristics} and \citet{thompson2003close}, who established many of the fundamental environmental indices used in modern tornado forecasting.

\subsection{SPC SFCOA dataset}

The GR-S HRRR-Inflow supercell dataset is compared to the widely used SPC convective mode dataset \citep[\citetalias{coffer2019srh500}]{smith2012modes,thompson2012convective}. Model-based environmental analyses for each event are derived from the SPC's Surface Objective Analysis \citep[SFCOA; ][]{bothwell2002integrated} underpinned by the RUC and RAP, with environmental fields interpolated from the native model vertical levels to pressure levels every 25 hPa. SFCOA merges surface observations with RUC/RAP analyses to obtain a more realistic representation of the near-storm environment. This is hereafter referred to as the ``SPC dataset'' or ``SPC SFCOA dataset''. The SPC dataset has been used in a plethora of studies to evaluate environments favorable for particular severe hazards \citep{thompson2012convective,anderson2016investigation,coffer2019srh500,warren2021spectrum,nixon2022distinguishing}, as well as to drive numerical modeling experiments \citep{goldacker2021updraft,goldacker2023srh,coffer2023LLM}. The generation of the storm events of the SPC dataset is described in detail by \citet{smith2012modes} and \cite{thompson2012convective}. The SPC dataset is the same right moving supercell events (discrete, cell-in-cluster, and cell-in-line supercells) used by \citetalias{coffer2019srh500} from 2005--2017, with an additional inclusion of tornadic supercells from 2018-2021. The nontornadic supercell dataset used by \citetalias{coffer2019srh500} is unchanged and includes the years 2005--2012 and 2014--2015 (after which nontornadic events were no longer being logged). The nontornadic supercells are primarily associated with significantly severe, right-moving supercells (i.e., at least 65 kts winds and/or 2-inch hail), outside of the years 2014--2015 where all severe thresholds were included in convective mode identification (i.e., at least 50 kts winds and/or 1-inch hail). This is in contrast to the GR-S HRRR-Inflow dataset, which is comprised of non-severe and severe supercells. For consistency with the GR-S HRRR-Inflow dataset, we again remove cases with zero MUCAPE, EIBs above the surface, or EBWD less than 12.5 m s$^{-1}$. In total, this eliminated 4051 cases ($~$17\$), leaving a total of 8637 nontornadic supercells, 9124 weakly tornadic supercells, and  1890 significantly tornadic supercells for further analysis. As will be shown in Section 3 (and as documented by \citetalias{coniglio2024sampling}), the general conclusions from the SPC dataset in \citetalias{coffer2019srh500} are unchanged. Finally, unlike the GR-S dataset, observed storm motions are not available for the SPC dataset and all storm-relative variables use the original formulation of the Bunkers storm motion \citep{bunkers2000predicting}. The effect of the Bunkers estimate on common forecasting indices will be evaluated in Section~\ref{GRSvsSPC-SM} using the GR-S dataset. 

\subsection{Analysis techniques}

Conventional operational indices are computed from vertical profiles of atmospheric state variables using the same independent calculation routine as in \citetalias{coffer2019srh500}, in order to expand the available parameters beyond what is routinely archived within the SPC mesoanalysis or the HRRR. Forecast skill for each variable across the full SPC and GR-S datasets is evaluated with the ``true-skill score'' (TSS; see Eq. 2 in \citetalias{coffer2019srh500}, also commonly know as the Pierce skill score), which is maximized when probability of detection (POD or ``hit rate'') is high while probability of false detection (POFD) is low \citep[i.e., TSS $=$ POD - POFD;][]{wilks2011statistical}. TSS for each variable is calculated at 100 discrete intervals\footnote{TSS was also calculated over 1000 discrete intervals for select variables to ensure the maximum skill was not underestimated, leading to an incorrect ranking of predictors. This test confirmed that 100 intervals was sufficient to accurately capture the maximum TSS.} between the 5th and 95th percentiles to determine the ``optimal threshold'' or TSS$_{max}$. A benefit of using TSS compared to some other measures of forecast skill is that it is independent of the event base rate (i.e., tornado occurrences), which allows comparisons across both the GR-S HRRR-Inflow and SPC datasets. Additionally, mean difference (MD; sometimes referred to as mean error or bias\footnote{The term `difference' is preferred over `error' or `bias' in this context as to not imply one model analysis is superior to the other, without corresponding observations.}), relative MD (i.e., a percentage difference relative to the mean value; RMD), and mean absolute difference (MAD) are used to compare environmental indices between SFCOA and HRRR-Inflow representations of the same events, with statistical significance tests computed via a Mann-Whitney U-test. 

\begin{figure*}[t]
\centerline{\includegraphics[width=40pc]{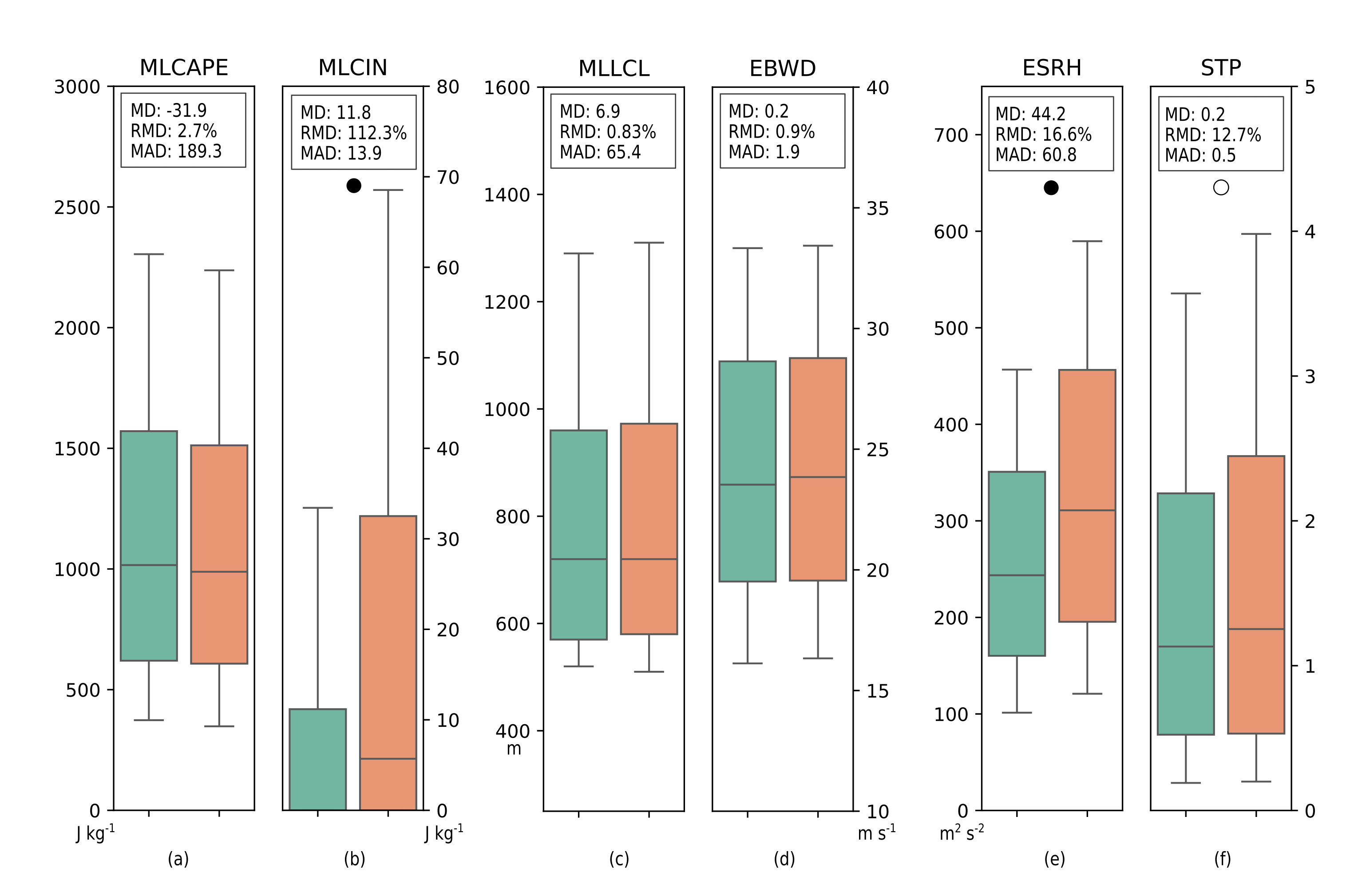}}
\caption{Box-and-whisker plots of each component of the effective-layer significant tornado parameter (STP) for the 2160 filtered, right-moving tornadic supercell events from the SPC dataset between 2017--2021 for both the SFCOA (green) and the HRRR-Inflow (orange) representations of the near-storm environment. The components are as follows: (a) mixed-layer convective available potential energy (MLCAPE), (b) mixed-layer convective inhibition (MLCIN), (c) mixed-layer lifting condensation level (MLLCL), (d) effective bulk wind difference (EBWD), and (e) effective storm-relative helicity (ESRH). (f) The resulting distributions of the STP composite index. The whiskers extend upward to the 90th and downward to the 10th percentiles. Outliers are excluded for clarity. For each plot, associated mean difference (MD), relative mean difference (RMD), and mean absolute difference (MAD), as described in Section~\ref{methods}c, are also listed. Distributions that are significantly different at a 90\% and 99\% confidence interval are indicated by an empty circle and filled circle, respectively.}
\label{fig:SFCOAvsHRRR_STPboxplot}
\end{figure*}

\section{Direct comparison between SFCOA and HRRR-Inflow}\label{HRRRvsSFCOA}


First we address whether the HRRR more closely mirrors field project datasets than previous model-based analyses due to its higher resolution and more sophisticated assimilation of radar-reflectivity and lightning observations. SFCOA analyses are compared  \emph{directly} to the HRRR-Inflow technique \emph{over the same events}. Spatially averaged inflow soundings from the HRRR were generated for all events in the SPC dataset during the years 2017--2021 (i.e., full years with overlapping availability of SPC cases and HRRRv2+). Cases were jointly filtered based on meeting the thresholds of MUCAPE, EBWD, and EIB height (discussed in Section~\ref{methods}). This resulted in 2160 tornadic supercell events to compare head-to-head, having removed any differences between construction of the supercell case list (i.e., SPC vs GR-S). Since the SPC dataset includes only tornadic events from 2015 onwards (and not nontornadic events), statistical analysis is limited to difference comparisons but not measures of forecast skill (i.e., skill scores, like TSS, require null events to evaluate a 2x2 contingency table). 

\begin{figure*}[t]
\centerline{\includegraphics[width=30pc]{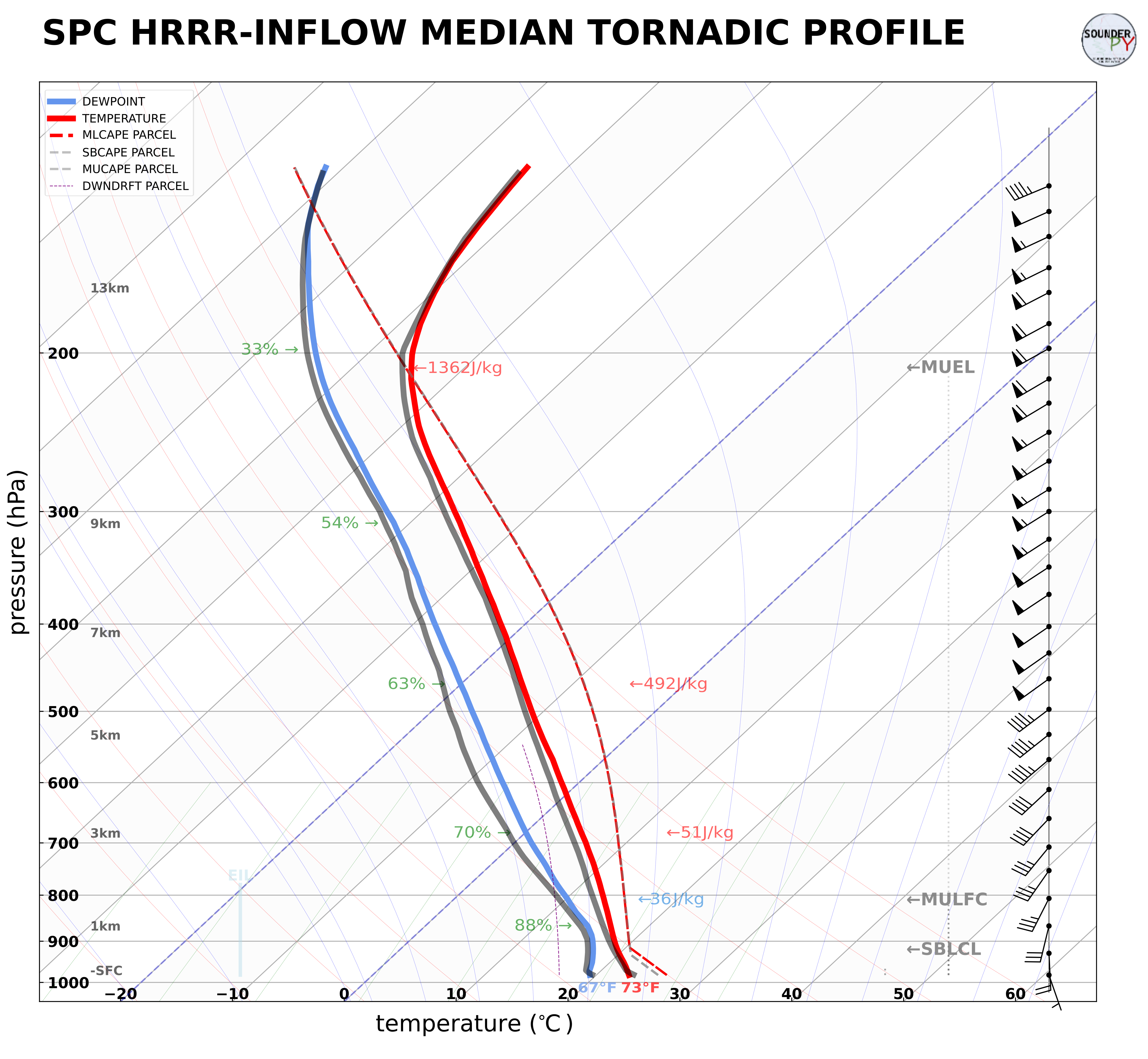}}
\caption{Composite skew$T$ log--$p$ diagrams for the 2160 tornadic supercells from the SPC dataset with environmental profiles extracted from the SFCOA (gray) and via the HRRR-Inflow technique (shaded). The parcel profile shown corresponds to the virtual temperature of the parcel.}
\label{fig:SFCOAvsHRRR_sounderpy}
\end{figure*}

\begin{figure*}[t]
\centerline{\includegraphics[width=25pc]{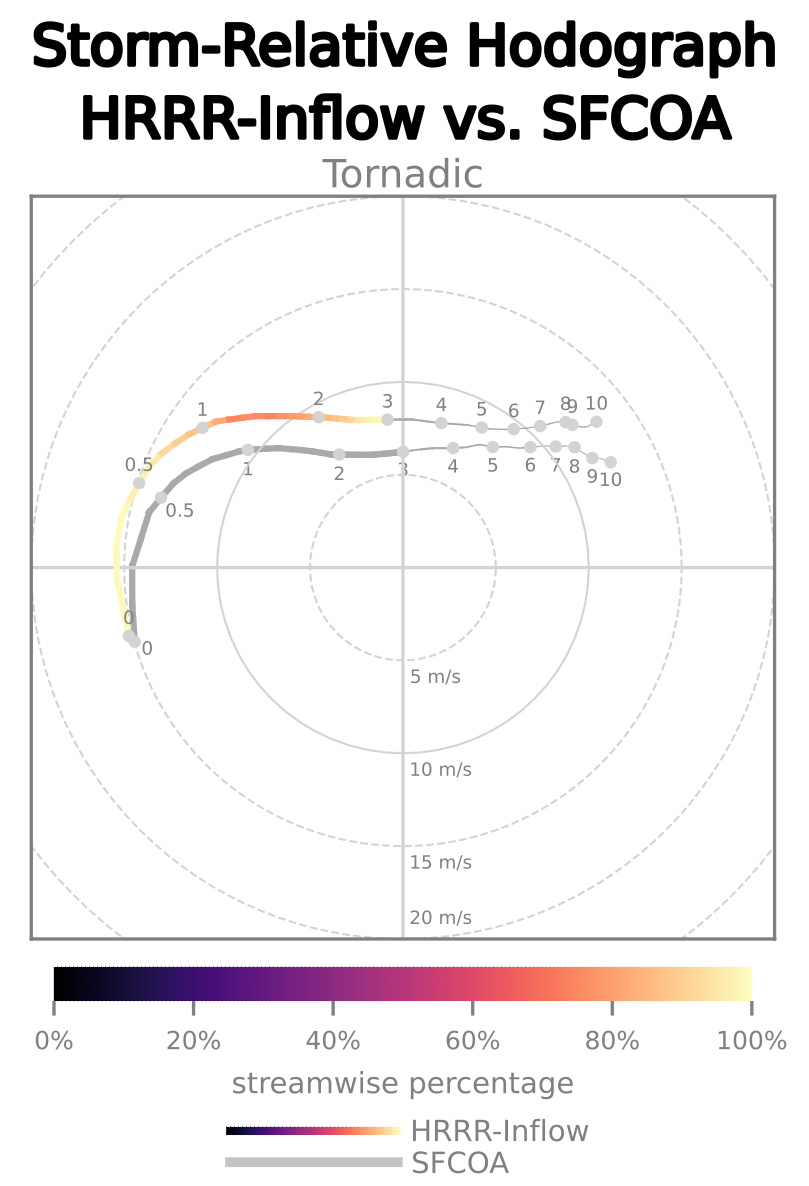}}
\caption{Composite storm-relative hodographs for the 2160 tornadic supercells from the SPC dataset with environmental wind profiles extracted from the SFCOA (gray) and via the HRRR-Inflow technique (shaded). The shaded values for the HRRR-Inflow hodographs, between the surface and 3 km AGL, represent the percent of total horizontal vorticity that is oriented in the streamwise direction calculated in 100 m increments. Height markers are indicated at 500 m AGL and then every kilometer starting at 1 km AGL. The storm motion, calculated via the Bunkers method for both datasets, is at the hodograph origin.}
\label{fig:SFCOAvsHRRR_storm_relative_hodographs}
\end{figure*}

\begin{figure*}[t]
\centerline{\includegraphics[width=40pc]{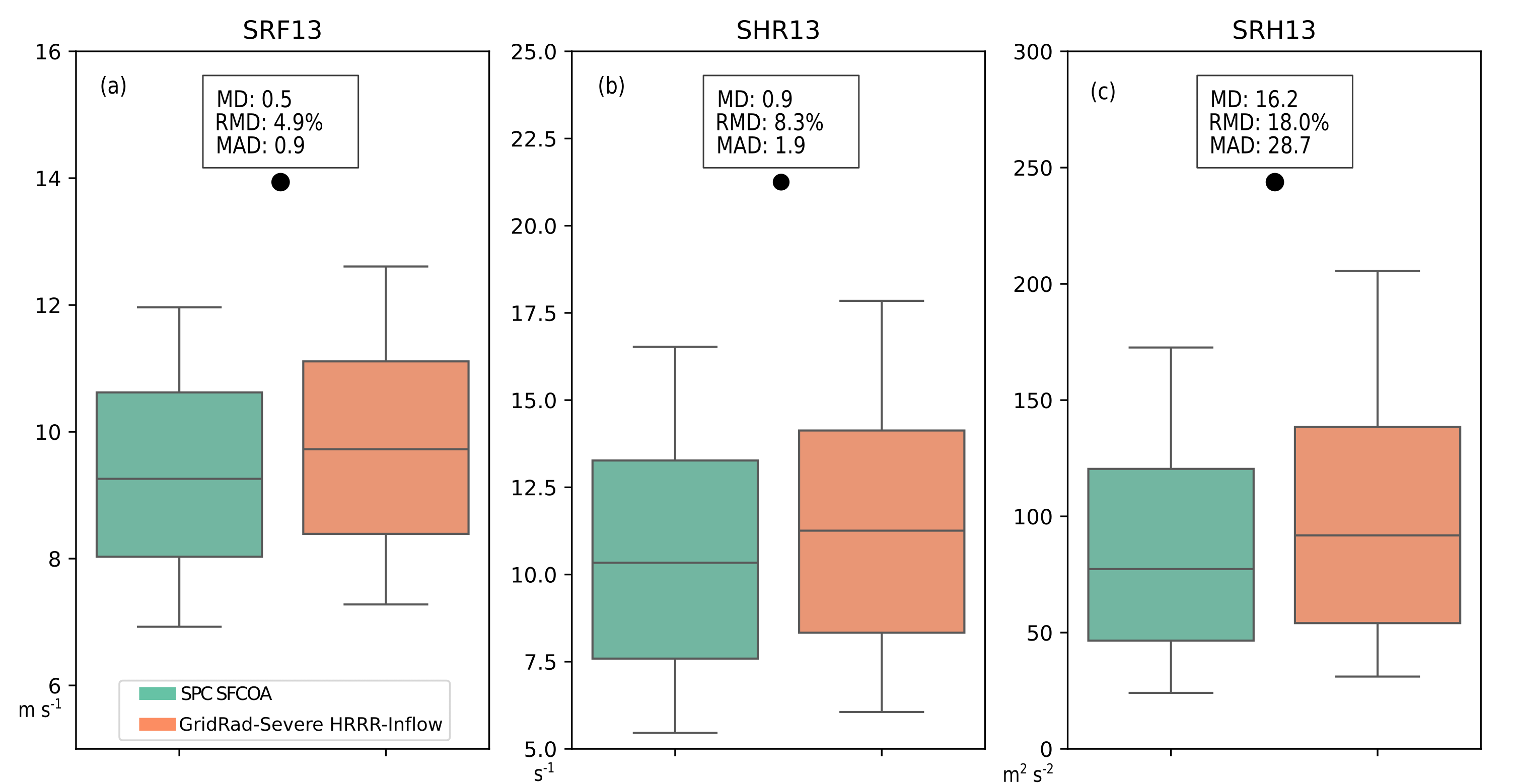}}
\caption{As in Figure~\ref{fig:SFCOAvsHRRR_STPboxplot}, except for the storm-relative flow (SRF), bulk vertical wind difference (BWD), and storm-relative helicity (SRH) in the 1--3 km AGL layer.}
\label{fig:SFCOAvsHRRR_MLWboxplot}
\end{figure*}

We start by comparing the various components of the effective-layer STP \citep{thompson2007effective}: CAPE, convective inhibition (CIN), and the lifting condensation level (LCL) for the mixed layer (ML) parcel, the effective bulk wind difference (EBWD), and ESRH. Distributions between the SFCOA and HRRR for MLCAPE are generally quite similar (Fig.~\ref{fig:SFCOAvsHRRR_STPboxplot}a), with an RMD under 3\%. MLCAPE is slightly lower in the HRRR compared to the SFCOA (MD = -32 J kg$^{-1}$), which is consistent with the small positive bias compared to observations in \citetalias{coniglio2022mesoanalysis}. Comparing lapse rates and humidity profiles throughout various layers indicates that the lower MLCAPE in the HRRR-Inflow is primarily due to warmer temperatures throughout the troposphere, evidenced by positive temperature differences in the composite skew$T$ (Fig.~\ref{fig:SFCOAvsHRRR_sounderpy}). This may be the result of storm influences on the environment from mid-level latent heating in the convection-allowing HRRR not present in a coarser model analysis  (which is partly captured due to the ``inflow'' averaging technique in the vicinity of model-represented convection). Related to the warmer temperature profiles, especially in the low-to-mid-levels, are positive MLCIN differences in the HRRR compared to the SFCOA (Fig.~\ref{fig:SFCOAvsHRRR_STPboxplot}b,\ref{fig:SFCOAvsHRRR_sounderpy}). The RMD is over 100\%, indicating stronger capping inversions in the HRRR. Although the magnitude of the RMD may seem quite large, the MD and MAD for MLCIN are both less than 15 J kg$^{-1}$ (Fig.~\ref{fig:SFCOAvsHRRR_STPboxplot}b) and the absolute values of CIN in these environments tends to be small ($<$ 30 J kg$^{-1}$). Despite the differences in the distributions being statistically significant (significance levels are shown with circles in Figs.~\ref{fig:SFCOAvsHRRR_STPboxplot},\ref{fig:SFCOAvsHRRR_MLWboxplot}), they are well within the range that would generally still be considered favorable for deep convection (i.e., the formulation of the STP does not differentiate between MLCIN values less than 50 J kg$^{-1}$). 
In similar comparisons between tornadic supercell soundings by \citetalias{coniglio2022mesoanalysis}, MLCIN was on average higher in the SFCOA than the observations. It therefore seems likely the HRRR does not represent an improvement over the SFCOA in this regard. While we suspect the warmer temperatures are due to convective heating represented in the HRRR, convection-allowing models are also known to have limited ability to predict the vertical structure of capping inversions \citep{burlingame2017pbl}. Finally, the distributions of MLLCL and EBWD between the two datasets are essentially identical with an RMD of less than 1\% (Fig.~\ref{fig:SFCOAvsHRRR_STPboxplot}c,d). Within the ML, the HRRR is generally warmer but moister (Fig.~\ref{fig:SFCOAvsHRRR_sounderpy}), yielding no significant differences in the LCL distributions (Fig.~\ref{fig:SFCOAvsHRRR_STPboxplot}c). As for EBWD, the shape of the composite hodographs between the SFCOA and the HRRR are relatively similar, especially above 3 km AGL (Fig.~\ref{fig:SFCOAvsHRRR_storm_relative_hodographs}). Again, this results in no statistically significant differences between the EBWD distributions (Fig.~\ref{fig:SFCOAvsHRRR_STPboxplot}d). 

In comparison to other common forecast indices discussed above, more definitive differences are seen in the lower tropospheric wind profile. ESRH is considerably higher in the HRRR compared to the SFCOA (Fig.~\ref{fig:SFCOAvsHRRR_STPboxplot}e), with an RMD of over 16\%. Statistically significant positive differences in SRH are present regardless of the depth of the integration layer (not shown). Although there is not a nontornadic dataset to evaluate measures of forecast skill, positive differences in SRH in the HRRR (compared to the SFCOA) are more pronounced within the significantly tornadic supercell subset than weakly tornadic supercells: the MD for ESRH increases from 38 m$^{2}$ s$^{-2}$ for weakly tornadic events to 71 m$^{2}$ s$^{-2}$ for significantly tornadic events. Across a wide range of studies, it has been shown that weakly tornadic supercell environments often share similarities with nontornadic supercell environments. Thus, it is reasonable to assume that the significantly tornadic supercells would have larger positive differences in SRH than nontornadic supercells (if they were available in the SPC dataset after 2015 for this specific head-to-head comparison).  
This will have ramifications when comparing measures of forecast skill between significantly tornadic and nontornadic supercells for kinematic environmental indices in Section~\ref{GRSvsSPC}b. 

Differences across the basic parameter space of MLCAPE, MLCIN, MLLCL, EBWD, and ESRH between the SFCOA and the HRRR might be expected to largely offset one another when combined into the STP index. Indeed, there is a rather small MD magnitude for STP (Fig.~\ref{fig:SFCOAvsHRRR_STPboxplot}f), which is almost entirely due to ESRH. Although the difference is marginally statistically significant, the MD of 0.2 for the STP is probably of no practical relevance, falling below the lowest contour interval on the SPC SFCOA mesoscale analysis webpage. Without corresponding radiosonde observations (RAOBs) for these events, we can not specifically answer the question of whether the HRRR provides an improved representation of tornadic storm environments compared to the SFCOA; however, the small differences for most variables and lack of statistically and practically significant differences (with the exception of MLCIN and ESRH) indicates that \emph{incorporating more frequent information on finer scales in the HRRR is not degrading the basic parameter space of common environmental indices from the original SFCOA analyses}, which was a concern in using a convection-allowing model--based analysis. Instead there appears to be a potential benefit of the HRRR (compared to \citetalias{coniglio2020insights} and \citetalias{coniglio2022mesoanalysis}), specifically related to the low-to-mid-level wind profile (reflected in ESRH). We next shift the focus to more specific details of the lower tropospheric wind profile and how the representation of the near-storm kinematic environment in HRRR compares to observational field projects versus prior model-based climatologies. 

Differences in SRH tend to accumulate with increasing depth, leading to the greatest differences between the SFCOA and the HRRR in the 0--3 km layer (SRH3). This is not simply due to evenly distributed differences throughout the lower troposphere though. The RMD for SRH increases from 6\% in the 0--500 m layer to almost 15\% in the 0--3 km layer (not shown). The hodograph representation of this increase in RMD for SRH can be seen in Figure~\ref{fig:SFCOAvsHRRR_storm_relative_hodographs}, where the distance between the SFCOA and HRRR composite hodographs increases with height, especially in the low-to-mid-level wind profile from 1--3 km AGL. The HRRR storm-relative hodograph displays more shear and faster storm relative flow in the low-to-mid-levels for the same events than the corresponding representation in the SFCOA (Fig.~\ref{fig:SFCOAvsHRRR_storm_relative_hodographs}).  Because the 1--3 km layer is emblematic of the biggest differences between the composite hodographs in Fig.~\ref{fig:SFCOAvsHRRR_storm_relative_hodographs}, as well as in \citetalias{coniglio2020insights}, this layer is a prominent focus for the rest of the present study. The distributions of storm-relative flow and vertical wind shear in the 1--3 km layer (SRF13 and BWD13, respectively) are also correspondingly higher in the HRRR compared to the SFCOA. While the MBs for these variables are quite modest, 0.5 m s$^{-1}$ for SRF13 and 0.9 s$^{-1}$ for BWD13 (Fig.~\ref{fig:SFCOAvsHRRR_MLWboxplot}a,b), the distributions are statistically different from each other, unlike many other common environmental indices discussed earlier. The combination of SRF13 and BWD13 increases yield an RMD value for SRH in the 1--3 km layer (SRH13) of 18\%. 

\begin{figure*}[t]
\centerline{\includegraphics[width=40pc]{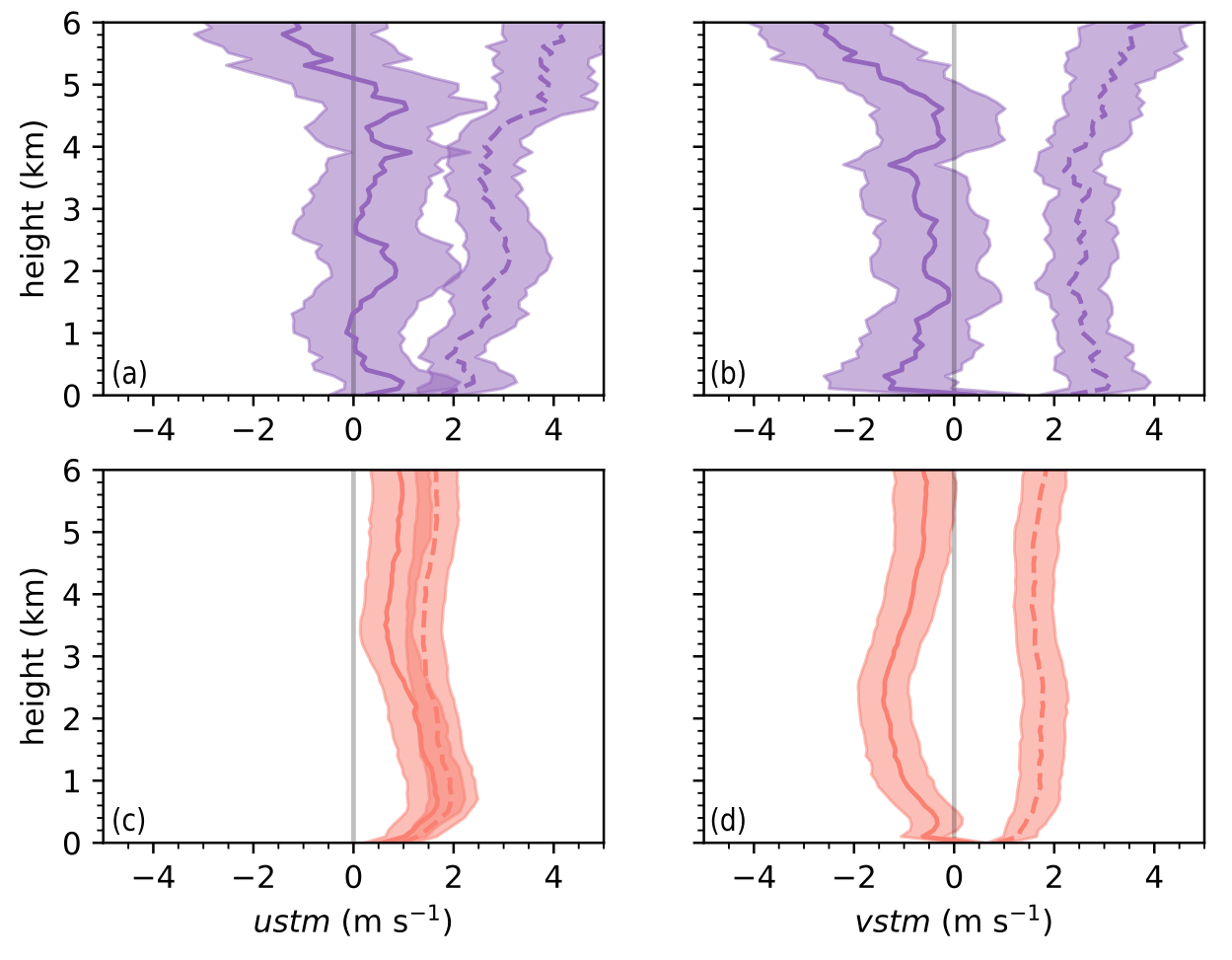}}
\caption{Difference profiles and 95\% confidence intervals of SFCOA mean difference (solid) and mean absolute difference (dashed) up to 6 km AGL of (left) storm-relative wind component along storm motion ($ustm$) and (right) storm-relative wind component normal to storm motion ($vstm$) derived from the (top; purple) `far-field TOR subset' profiles (n=62) in \citet{coniglio2022mesoanalysis} and (bottom; red) the HRRR-Inflow profiles of filtered, right-moving tornadic supercell events from SPC dataset between 2017--2021 (n=2190). As-in \citetalias{coniglio2022mesoanalysis}, for each wind profile, the x-axis is rotated to align with the storm motion before all calculations are performed; thus, $ustm$ and $vstm$ are understood to be components along and across the storm motion, respectively, and differ slightly from traditionally defined storm-relative wind components throughout the rest of the article.}
\label{fig:SFCOAvsHRRR_coniglio}
\end{figure*}

Although the mean differences between the HRRR and SFCOA within the low-to-mid-level wind profiles are relatively small in magnitude (despite the statistical significance), the composite hodograph differences between HRRR and SFCOA (Fig.~\ref{fig:SFCOAvsHRRR_storm_relative_hodographs}) mirror those between observed RAOBs and the SFCOA found by \citetalias{coniglio2022mesoanalysis}. In other words, the HRRR appears more similar to RAOBs than the SFCOA. Just as in Figure~\ref{fig:SFCOAvsHRRR_storm_relative_hodographs}, observed soundings in \citetalias{coniglio2022mesoanalysis} have larger shear magnitudes and faster storm relative flow below 3 km AGL than the same representation in the SFCOA (see their Figs. 6, 11). This is further supported by vertical profiles of MD and MAD of the storm-relative wind components between the HRRR and SFCOA from the present study compared to those between RAOBs and the SFCOA from \citetalias{coniglio2022mesoanalysis} (see their Figs. 3,7 and reproduced for completeness for their far-field tornadic cases in Figure~\ref{fig:SFCOAvsHRRR_coniglio}a,b)\footnote{There was insufficient number of cases (8) in the \citetalias{coniglio2022mesoanalysis} dataset since the implementation of HRRRv2 in late 2016 to directly compare the HRRR to RAOBs for near-storm supercell environments.}. Both storm relative wind components, $ustm$ and $vstm$,  in the HRRR vs. SFCOA from the present study (Fig.~\ref{fig:SFCOAvsHRRR_coniglio}c,d) have mean differences that are mostly the same sign compared to RAOBs vs. SFCOA from \citetalias{coniglio2022mesoanalysis} (Fig.~\ref{fig:SFCOAvsHRRR_coniglio}a,b), albeit the differences are generally larger in magnitudes between 1 and 3 km AGL herein. Overall, $vstm$ generally displays a negative difference and $ustm$ displays a positive difference below 3 km (Fig.~\ref{fig:SFCOAvsHRRR_coniglio}), again indicating that the HRRR is more similar to the RAOBs than the SFCOA. \citetalias{coniglio2022mesoanalysis} argued that SFCOA underestimates the low-level wind shear, especially near the surface, due to a lack of vertical resolution near the ground and errors linked to the planetary boundary layer parameterization. The HRRR-Inflow dataset seems to at least partly compensate for this (Figs.~\ref{fig:SFCOAvsHRRR_storm_relative_hodographs},\ref{fig:SFCOAvsHRRR_MLWboxplot},\ref{fig:SFCOAvsHRRR_coniglio}). 
These differences in the storm-relative wind profile mostly explain the positive differences in SRH throughout the lower troposphere in the HRRR compared to the SFCOA shown in Figure~\ref{fig:SFCOAvsHRRR_MLWboxplot}. 



In conclusion, when it comes to the storm-relative wind profile, \emph{the HRRR-Inflow appears more similar to observational field project datasets than the same representation in the SFCOA}, while other common forecast indices are relatively similar between the two model analyses. We presented these direct comparisons (across common cases) between the HRRR and SFCOA upfront to show that the HRRR can appropriately represent some of the features that appear to be missing in the SFCOA. 
We next ask whether an independent dataset of supercells using the GR-S and the HRRR can help further address discrepancies in prior work. 

\section{GR-S HRRR-Inflow vs SPC SFCOA datasets} \label{GRSvsSPC}

In the previous section, we compared the representation of near-storm environments across two different model analyses, the SFCOA and the HRRR-Inflow, over the same set of tornadic events in the SPC convective mode dataset. This direct comparison removed any potential differences between the supercell event datasets (i.e., SPC vs GR-S), but was limited to only tornadic events since the SPC dataset ceased to log nontornadic events after 2015, and thus, prior to the modern HRRR era. We next seek to understand if the environmental differences between the HRRR-Inflow and SFCOA analyses noted in the head-to-head comparisons from Section~\ref{HRRRvsSFCOA}, such as in the low-to-mid-level wind profile, hold across a larger catalog of cases in the GR-S supercell dataset for both nontornadic and tornadic supercells (and how this compares to the full SPC SFCOA dataset). If similar trends in the low-to-mid-level wind profile exist, this would begin to bridge the gap between \citetalias{coffer2019srh500} and \citetalias{coniglio2020insights}.   
 
\subsection{Geographic and temporal comparison}

Before evaluating the forecast skill of individual environmental variables in the GR-S dataset, it is worth first exploring inherent spatial and temporal differences between the SPC and GR-S datasets that arise due to their differing methods of construction. Geographic and temporal biases have the potential to make certain variables appear to have more or less forecast skill \citepalias[e.g.,][]{coniglio2024sampling}.  While there are clear advantages to using GR-S to generate an independent dataset of supercells (with observed storm motions), does the GR-S dataset represent climatology as well as previous, large event population studies did? 

\begin{figure*}[t]
\centerline{\includegraphics[width=40pc]{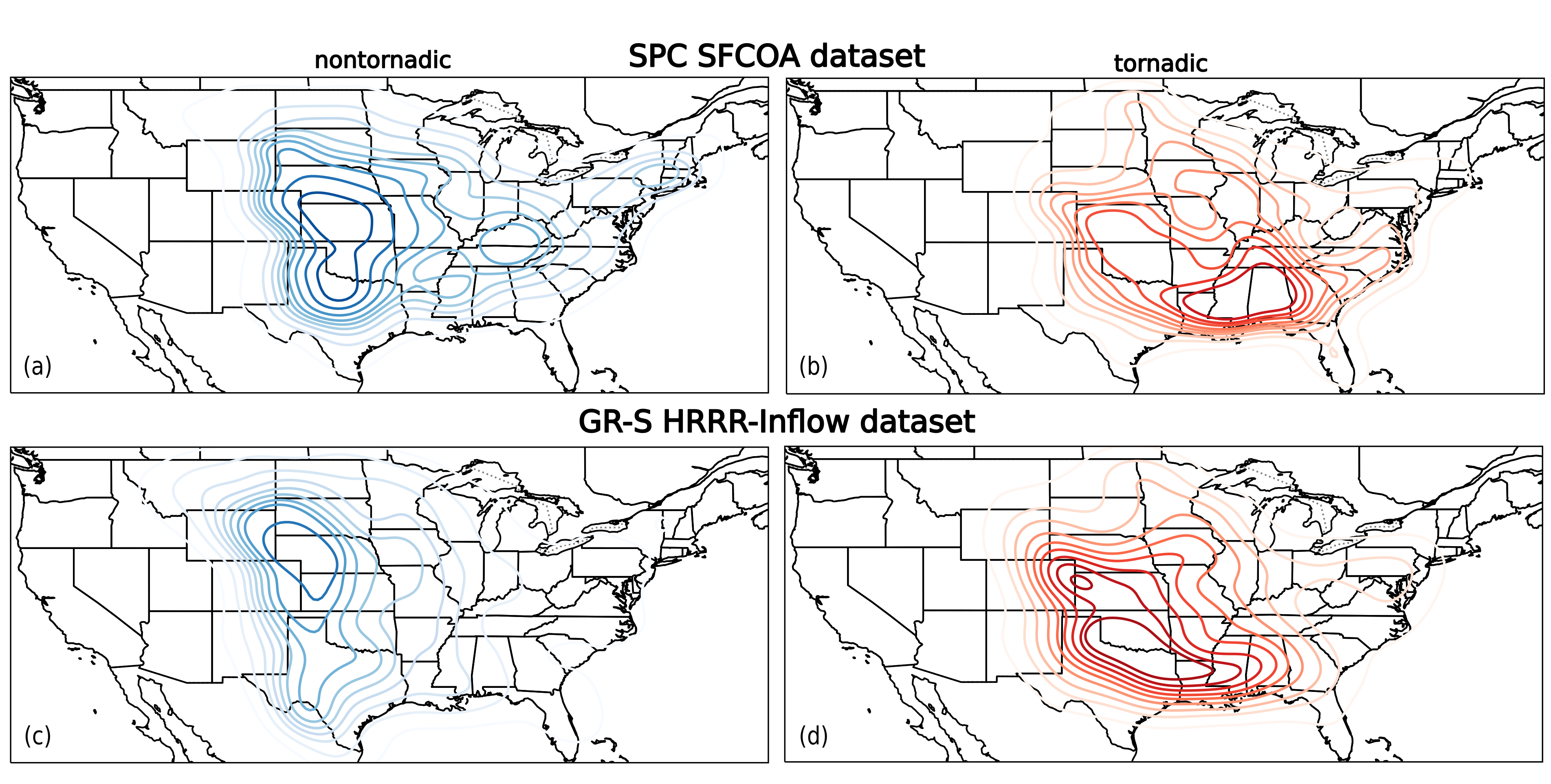}}
\caption{Kernel density estimate (KDE) of reports over the entire datasets in (a,b) the SPC SFCOA and (c,d) GR-S HRRR-Inflow datasets for nontornadic (blue; a,c) and tornadic (red; b,d) events. KDE contours for the respective datasets are centered on the point of highest event density and contain contours that represent 10, 20, 30, 40, 50, 60, 70, 80, and 90\% of the event data, where darker contours represent the highest event density.}
\label{fig:reportKDEs}
\end{figure*}

Supercell tornado occurrence in the SPC dataset maximizes from the central Great Plains southeastward to the lower Mississippi Valley, with the highest occurrences stretching from Louisiana, Mississippi, and into Alabama (Fig.~\ref{fig:reportKDEs}b, and similar to Fig. 6 in Smith et al. 2012). This extension into the southeastern United States over the last 20 years is consistent with shifts in long-term tornado trends \citep{gensini2018spatial}. In contrast, nontornadic supercell occurrence within the SPC dataset peaks in the Great Plains of Kansas and Oklahoma (Fig.~\ref{fig:reportKDEs}a). In previous studies this has been partially attributed to the filtering of potential events in the SPC dataset being linked to significant severe hail and wind reports (except in 2014-15). This results in rather distinct spatial areas of tornadic and nontornadic supercell events within the SPC dataset. The displacement affects the skill of certain environmental variables, such as the potential for tornadic events to have lower CAPE and higher SRH due to regional effects and seasonal effects (discussed more in \citetalias{coniglio2024sampling}). 

Nontornadic and tornadic supercell events in the GR-S dataset are more spatially coincident with one another than in the SPC dataset (Fig.~\ref{fig:reportKDEs}c,d), although the highest densities are still distinct latitudinally across the Great Plains. In GR-S, tornadic events are most common through Kansas, Oklahoma, and into northern Texas (Fig.~\ref{fig:reportKDEs}d), while nontornadic events peak in the High Plains, with the highest density of events in the Nebraska Panhandle, South Dakota, and eastern Wyoming (Fig.~\ref{fig:reportKDEs}c). In particular, there are noticeably few nontornadic supercells in the southeastern United States in GR-S. The distribution of nontornadic events in GR-S is similar to the distribution of overshooting tops using full GridRad climatology \citep{cooney2018ten}, which may speak to how storm cells are identified in GR-S. 

\begin{figure*}[t]
\centerline{\includegraphics[width=30pc]{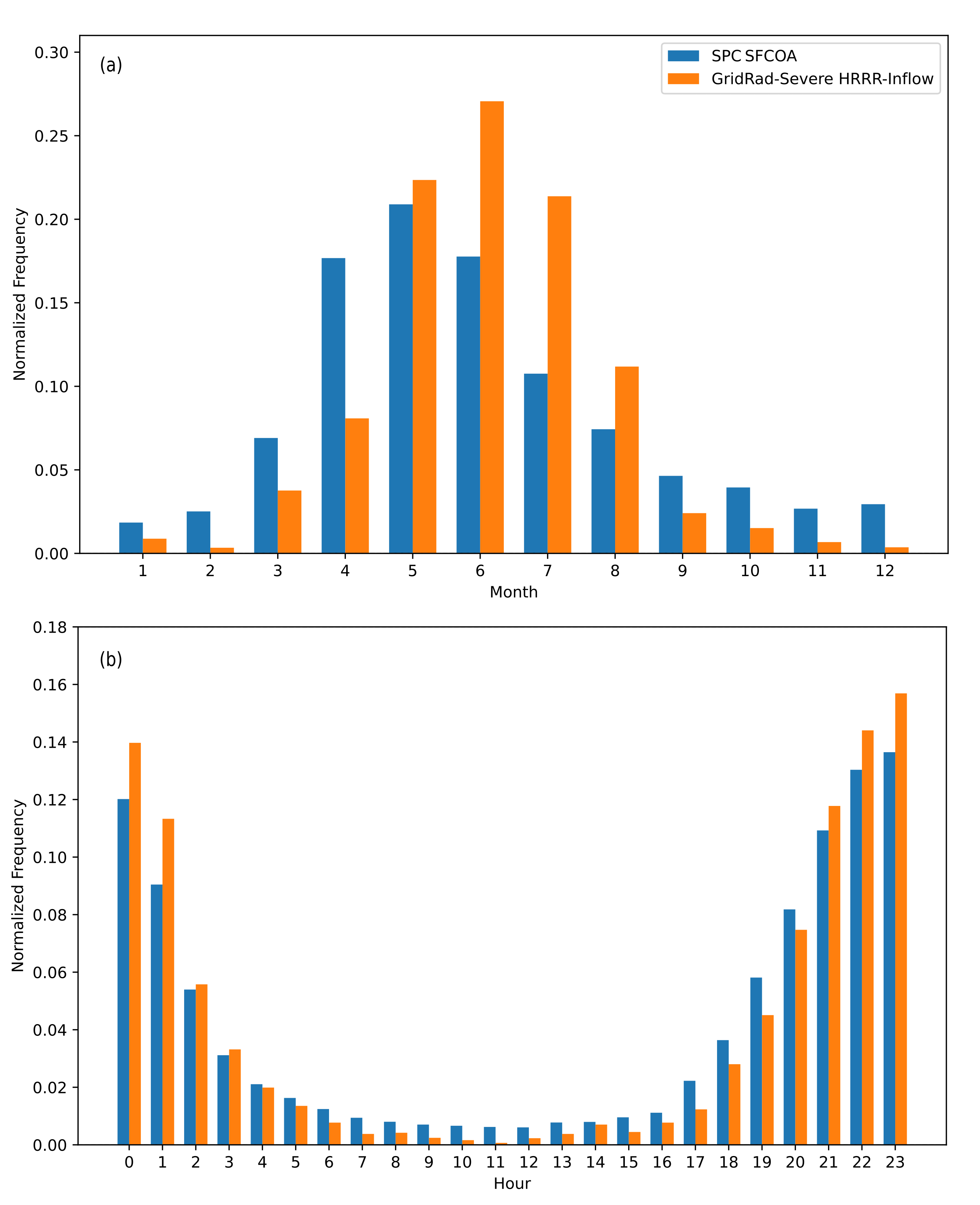}}
\caption{Normalized histograms of the frequency of months (top) and hour (bottom) from the SPC SFCOA dataset (blue) and the GR-S HRRR-Inflow dataset (orange).}
\label{fig:reportTemporal}
\end{figure*}

There are also temporal differences between the two datasets that could affect the interpretation of the subsequent results. Generally, the GR-S dataset has more cases during warmer months of the year and hours of the day (Fig.~\ref{fig:reportTemporal}). In the SPC dataset, supercells most often occur from April to June with a steady number of cases from September to February (Fig.~\ref{fig:reportTemporal}a). The highest occurrence of supercells throughout the day in the SPC dataset is from 20 to 02Z (Fig.~\ref{fig:reportTemporal}b). Both of these trends mirror the observed severe weather climatology in storm event data \citep{NOAA2022}. On the other hand, the GR-S dataset peaks a month later, from May to July (Fig.~\ref{fig:reportTemporal}a), and has noticeably fewer cases, in terms of normalized frequency, during the winter months and overnight hours compared to the SPC dataset (Fig.~\ref{fig:reportTemporal}b). The bias in GR-S events towards warmer months of the year and times of day is primarily driven by nontornadic events, as the temporal climatology of significantly tornadic events is quite similar between the GR-S and SPC datasets (not shown). This nontornadic bias may be due to a higher 30-dBZ echo top height threshold for nontornadic events than tornadic events (6 vs 4 km AMSL)\footnote{As a reminder, the differing height thresholds was a compromise to maximize the sample size of tornadic events, especially in the cool season and southeast United States, while keeping the nontornadic events to a reasonable sample size. Because storm tracks in GR-S can be for non-severe and severe convection, lowering the height threshold to 4 km for nontornadic events resulted in an exponential increase in cases, most of which were not of particular relevance. Restricting the tornadic subset to echo top heights greater than 6 km did not drastically change either thermodynamic nor kinematic composite profiles.}. 

To recap, all objective sample construction methods have drawbacks. A weakness of the SPC dataset is that nontornadic supercells are almost all significantly severe, which implicitly self-selects nontornadic events from a smaller part of the supercell parameter space (adding regional and seasonal biases not seen in the tornadic events); on the other hand, a weakness of GR-S is that pre-determined thresholds are applied unilaterally, which may also induce temporal and spatial biases that manifest themselves differently than in the SPC. We acknowledge upfront these potential biases in the construction of the GR-S dataset that could influence the trends in the apparent forecast skill of various convective indices, such as fewer cases in the cool season and in the southeastern United States; however, the GR-S dataset used in the present study appears as balanced, regionally and temporally, as canonical tornado environmental studies, such as \citet{thompson2003close}. 
The benefits of using GR-S, especially the ability to objectively generate a large, independent dataset of nontornadic and tornadic supercells in the modern HRRR era (with observed storm motions), make it ideal for addressing the research questions posed in this study. 

\begin{table*}
\caption{Maximum TSS (TSS$_{max}$) and corresponding value of the variable at the TSS maximum (VAR$_{max}$) at the optimal threshold for various forecasting parameters for discriminating between significant tornadic supercells (EF2+) and nontornadic supercells is given for the SPC SFCOA dataset and the GridRad-Severe HRRR-Inflow dataset. For the latter, further delineation is given between Bunkers storm motion (SM) estimates and the actual SM from GR-S. For variables that are not storm-relative, TSS$_{max}$ and VAR$_{max}$ are omitted and represented by a dash. The 10 variables with the highest forecast skill for each dataset and definition of storm motion are indicated and ranked accordingly in bold.}
\label{table:TSSmax}
\centering
\begin{tblr}{
  cells = {c},
  cell{1}{1} = {r=3}{},
  cell{1}{2} = {c=2}{},
  cell{1}{4} = {c=4}{},
  cell{2}{2} = {c=2}{},
  cell{2}{4} = {c=2}{},
  cell{2}{6} = {c=2}{},
  vline{2-3,8} = {1}{},
  vline{3,5,7} = {2}{},
  vline{4,6,8} = {3}{},
  vline{4} = {1--29}{},
  vline{2,6,8} = {2-29}{},
  hline{2-3} = {2-7}{},
  hline{4,26,30} = {-}{},
}
Variables & SPC SFCOA &        & GridRad-Severe HRRR-Inflow &        &            &        \\
          & Bunkers SM               &        & Bunkers SM                         &        & Actual SM        &        \\
          & TSS$_{max}$            & VAR$_{max}$ & TSS$_{max}$                      & VAR$_{max}$ & TSS$_{max}$     & VAR$_{max}$ \\
GRW500    & 0.49 \textbf{(8)}          & 9.9    & 0.495 \textbf{(10)}                  & 9.6   & -- \textbf{(10)}  & --     \\
GRW1      & 0.553 \textbf{(5)}         & 12.1   & 0.590 \textbf{(9)}                   & 10.3   & --         & --     \\
GRW6      & 0.549 \textbf{(6)}         & 19.1   & 0.692 \textbf{(1)}                   & 16.5   & -- \textbf{(1)}     & --     \\
BWD500    & 0.53 \textbf{(7)}          & 8.5    & 0.609 \textbf{(8)}                   & 6.8    & -- \textbf{(9)}     & --     \\
BWD1      & 0.554 \textbf{(4)}         & 11.9   & 0.635 \textbf{(5)}                   & 10.7   & -- \textbf{(5)}     & --     \\
BWD3      & 0.45              & 19.47  & 0.624 \textbf{(7)}                   & 16.9   & -- \textbf{(6)}     & --     \\
BWD13      & 0.02              & 12.8  & 0.234                   & 10.8   & --      & --     \\
BWD6      & 0.358             & 26.8   & 0.481                       & 24.4   & --         & --     \\
EBWD      & 0.357             & 25.9   & 0.486                       & 23.9   & --         & --     \\
SRF500    & 0.367             & 16.3   & 0.451                       & 14.7   & 0.27       & 13.9   \\
SRF1      & 0.279             & 14.9   & 0.426                       & 15.1   & 0.263      & 14.2   \\
SRF3      & 0.07              & 10.2   & 0.244                       & 10.9     & 0.285      & 10.1   \\
SRF13      & -0.01              & 12.5   & 0.174                       & 9.6     & 0.356      & 10.6   \\
SRH500    & 0.565 \textbf{(1)}         & 133.3  & 0.672 \textbf{(2)}                   & 105.2  & 0.619 \textbf{(7)} & 96.67  \\
SRH1      & 0.556 \textbf{(3)}         & 182.8  & 0.661 \textbf{(3)}                   & 143.8  & 0.646 \textbf{(3)}  & 127.5  \\
SRH3      & 0.484 \textbf{(9)}        & 275.4  & 0.634 \textbf{(6)}                  & 253.3  & 0.674 \textbf{(2)}  & 235.2  \\
SRH13      & -0.01         & 8.4  & 0.210                   & 111.5  & 0.451   & 77.7  \\
ESRH      & 0.471 \textbf{(10)}        & 260.3  & 0.618 \textbf{(7)}                   & 235.7  & 0.641 \textbf{(4)}  & 254.4  \\
SW500     & 0.564 \textbf{(2)}         & 0.017  & 0.650 \textbf{(4)}                   & 0.013  & 0.618 \textbf{(8)}  & 0.015  \\
CW500     & 0.018             & 0.001  & 0.010                       & 0.014  & 0.03          & 0.0003      \\
PSW500    & 0.28              & 60.1   & 0.374                         & 55.5   & 0.330       & 54.5   \\
CA        & 0.27              & 83.3   & 0.377                       & 93.3   & 0.219      & 105.7  \\
MLCAPE    & -0.09             & 900.34   & 0.093                       & 1671.3  & --         & --     \\
ECAPE     & -0.059             & 847.7 & 0.07                       & 587.2  & 0.1        & 871.2  \\
MLCIN     & 0.156             & 0      & 0.242                       & 16.0    & --         & --     \\
MLLCL     & 0.362             & 890.9  & 0.423                       & 1040.7  & --         & --     \\
SCP       & 0.316             & 7.6    & 0.458                       & 6.7    & 0.561      & 6.1    \\
STP       & 0.368             & 1.2    & 0.528                       & 1.00   & 0.575      & 0.81   \\
STP500    & 0.473             & 0.85   & 0.675                       & 0.92   & 0.610      & 0.748  
\end{tblr}
\end{table*}

\subsection{Comparison of environmental indices}

\begin{figure*}[t]
\centerline{\includegraphics[width=40pc]{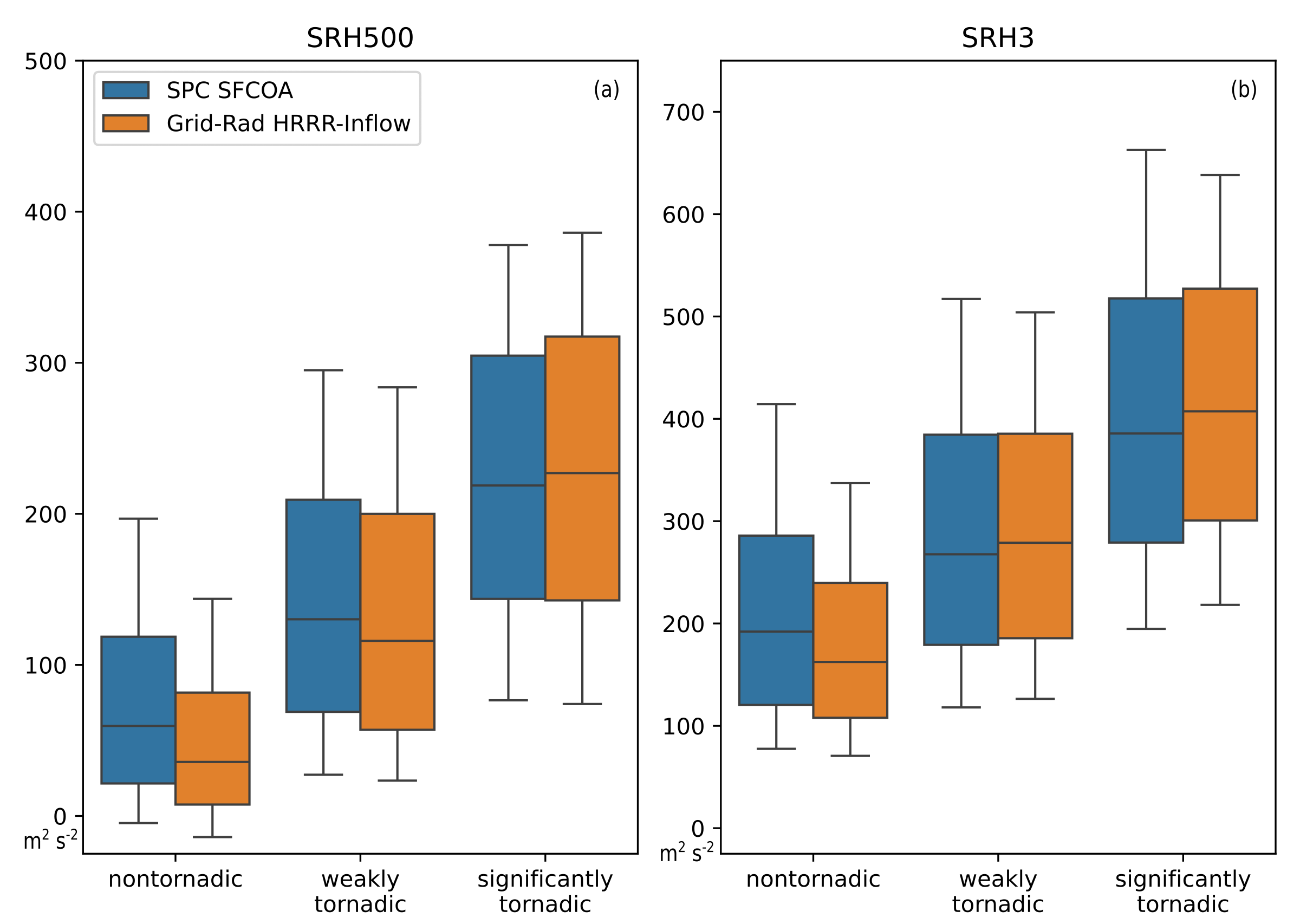}}
\caption{Box-and-whisker plots of storm-relative helicity (SRH) in the 0--500 m and 0--3 km AGL layers for all supercell events, separated by nontornadic supercells, weak tornadoes [(E)F0–1], and significant tornadoes [(E)F2+] from both the SPC SFCOA dataset (blue) and the GR-S HRRR-Inflow dataset (orange). The whiskers extend upward to the 90th and downward to the 10th percentiles. Outliers are excluded for clarity.}
\label{fig:SRHboxplot}
\end{figure*}
\begin{figure*}[t]
\centerline{\includegraphics[width=40pc]{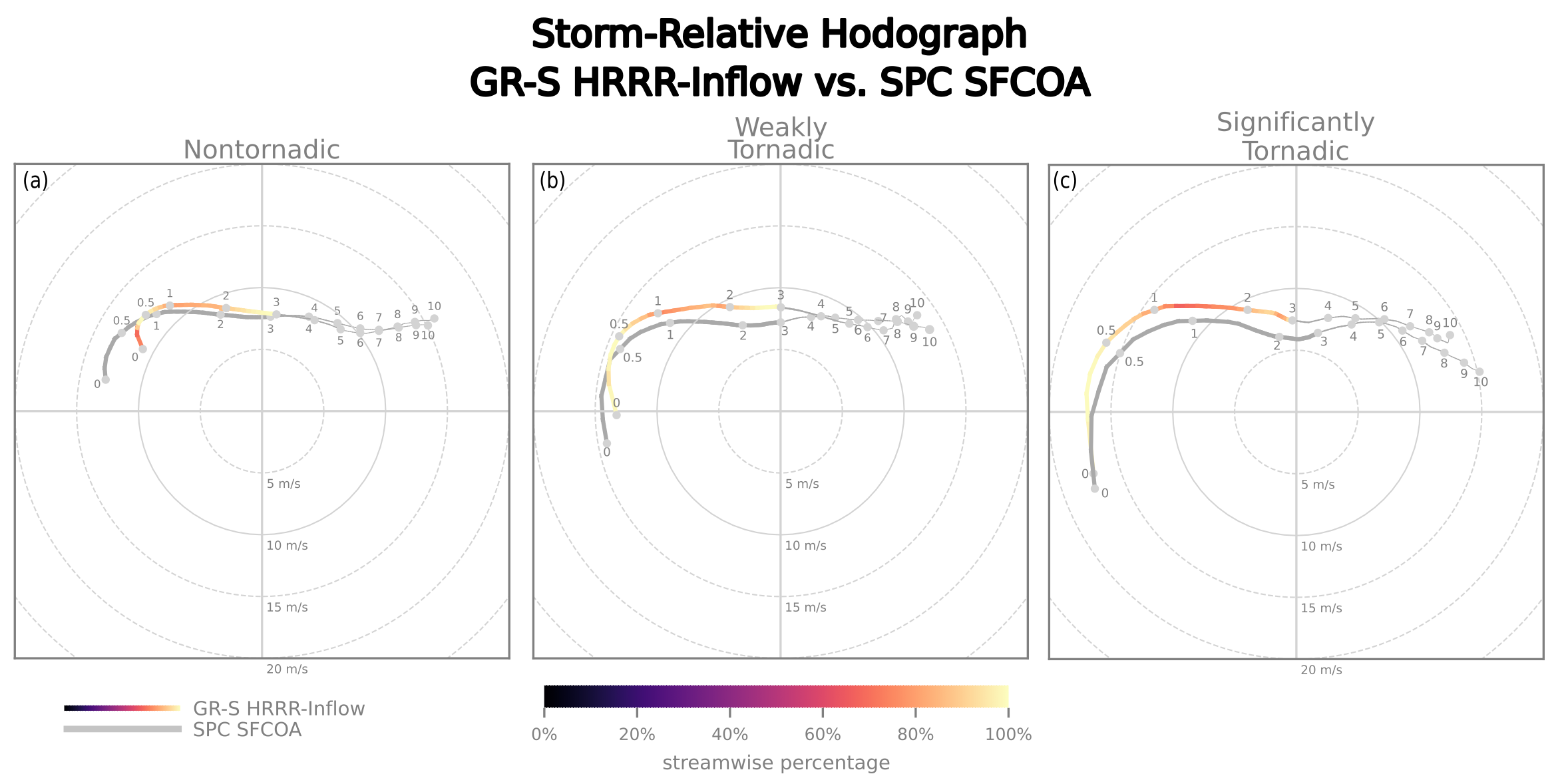}}
\caption{As in Figure~\ref{fig:SFCOAvsHRRR_storm_relative_hodographs}, except for (a) nontornadic, (b) weakly tornadic, and (c) significantly tornadic supercells from both the SPC SFCOA dataset (gray; up to 7 km AGL) and the GR-S HRRR-Inflow dataset (shaded). The storm motion, calculated via Bunkers for both datasets, is at the hodograph origin.}
\label{fig:storm_relative_hodographs}
\end{figure*}
\begin{figure*}[t]
\centerline{\includegraphics[width=40pc]{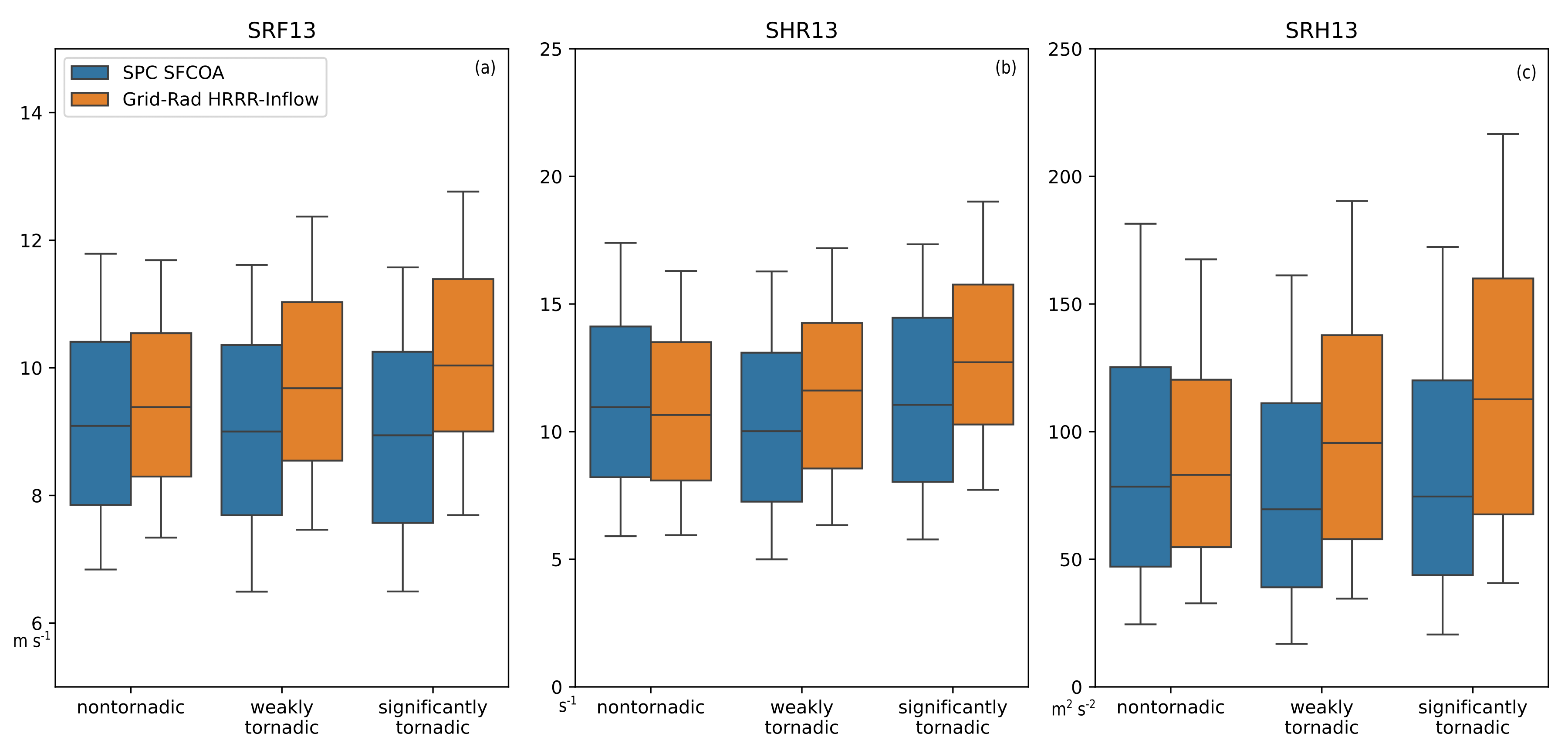}}
\caption{As in Figure~\ref{fig:SRHboxplot}, except for the storm-relative flow (SRF), bulk vertical wind difference (BWD), and storm-relative helicity (SRH) in the 1--3 km AGL layer.}
\label{fig:MLWboxplot}
\end{figure*}

To further address whether GR-S HRRR-Inflow supercell dataset displays characteristics that more closely mirror field project datasets rather than prior model-based tornado climatologies, we next explore composite hodographs, distributions of common environmental indices, and bulk TSS$_{max}$ statistics within the GR-S HRRR-Inflow dataset, while referring back to the full SPC SFCOA dataset as a baseline. The comparisons below primarily contrast nontornadic versus significantly tornadic supercells\footnote{Comparisons between nontornadic and all tornadic supercells revealed similar trends noted below (albeit lower values of TSS$_{max}$) and in \citetalias{coniglio2024sampling}. Although not explicitly shown from GR-S HRRR-Inflow dataset, those similar trends are implied by Figures~\ref{fig:SRHboxplot},\ref{fig:storm_relative_hodographs},\ref{fig:MLWboxplot}.} using Bunkers storm motion (the impact of observed storm motions will be evaluated in Section~\ref{GRSvsSPC-SM}). Although many different environmental indices were tested (some of these are listed in Table~\ref{table:TSSmax}), we primarily focus the following discussion on the wind profile as this is the main difference between \citetalias{coffer2019srh500} and \citetalias{coniglio2020insights}, as well as in the head-to-head comparisons between the SFCOA and HRRR in Section~\ref{HRRRvsSFCOA}.

As in most previous model-based environmental studies, TSS$_{max}$ increases for shallower layers of SRH, with SRH500 displaying the highest skill of any layer computed in the GR-S HRRR-Inflow dataset (Table~\ref{table:TSSmax}). The overall TSS$_{max}$ values for any layer of SRH is also higher in the GR-S HRRR-Inflow than in the SPC dataset (i.e., TSS$_{SRH500}$ = 0.672 vs. 0.565), indicating a greater discrimination between nontornadic and significantly tornadic supercells; however, the gains in forecast skill by going to these progressively shallower layers are lower in the GR-S HRRR-Inflow dataset (Table~\ref{table:TSSmax}). In other words, the relative increase in skill when decreasing the depth of integration of SRH from 3 km to 500 m is over 50\% smaller in GR-S HRRR-Inflow ($\Delta$TSS$_{max}$ = $+$0.038) compared to the same layers in the SPC SFCOA dataset ($\Delta$TSS$_{max}$ = $+$0.081). This results in only 4\% more correctly identified cases using SRH500 versus SRH3 in the GR-S HRRR-Inflow dataset compared to 10\% in the SPC SFCOA dataset. 
The changes in relative forecast skill in GR-S HRRR-Inflow are evident by comparing distributions of SRH500 versus SRH3 (Fig.~\ref{fig:SRHboxplot}) and composite hodographs (Fig.~\ref{fig:storm_relative_hodographs}) between nontornadic and tornadic supercells across datasets. For nontornadic supercells, the magnitude of SRH500 in the GR-S HRRR-Inflow dataset is 30-50 m$^2$ s$^{-2}$ lower than in the SPC SFCOA dataset, with the entire GR-S HRRR-Inflow distribution noticeably shifted to lower values (Fig.~\ref{fig:SRHboxplot}a). On average, for nontornadic supercells, the near-ground wind profile in the GRS-S HRRR-Inflow dataset has slower storm-relative winds and less streamwise vorticity in the lowest 500 m AGL than in the SPC SFCOA (Fig.~\ref{fig:storm_relative_hodographs}a). Composite wind profiles below 500 m are much more alike for tornadic supercells (Fig.~\ref{fig:storm_relative_hodographs}b,c), yielding relatively similar distributions of SRH500 between SPC SFCOA and GR-S HRRR-Inflow datasets (Fig.~\ref{fig:SRHboxplot}a). Above 500 m AGL, differences in the wind profile for nontornadic supercells between the GR-S HRRR-Inflow and the SPC SFCOA are essentially nonexistent, with the composite hodographs practically overlapping with each other (Fig.~\ref{fig:storm_relative_hodographs}a). Even so, tornadic supercells (and especially significantly tornadic supercells) display stronger storm relative winds and larger shear magnitudes above 500 m in the GR-S HRRR-Inflow (Fig.~\ref{fig:storm_relative_hodographs}b,c), yielding greater values of SRH3 (Fig.~\ref{fig:SRHboxplot}b). 

We showed in Section~\ref{HRRRvsSFCOA} that the low-to-mid-level wind profile from 1 to 3 km AGL is the most evident disparity between the HRRR and SFCOA over the same supercell events. This trend continues within the full dataset comparison where, unlike in any other model-based tornado climatology study to date (such as those that employ the SPC SFCOA dataset), there are meaningful differences in the low-to-mid-level wind profile between nontornadic and significantly tornadic supercell cases in the GR-S HRRR-Inflow (Figs.~\ref{fig:storm_relative_hodographs},\ref{fig:MLWboxplot}). As in the observed soundings datasets of \citetalias{coniglio2020insights} and \citetalias{coniglio2024sampling}, storm-relative winds in the 1--3 km AGL layer are on average a 1--2 m s$^{-1}$ faster in significantly tornadic supercells compared to nontornadic supercells in the GR-S HRRR-Inflow dataset (Fig.~\ref{fig:MLWboxplot}a). 
The distributions of 1--3 km bulk vertical wind shear for nontornadic supercells are very similar between the datasets but are again 1--2 m s$^{-1}$ larger for tornadic supercells in the GR-S HRRR-Inflow (Figs.~\ref{fig:MLWboxplot}b). The combination of faster storm-relative winds and larger bulk shear values result in SRH13 being substantially more skillful in the GR-S HRRR-Inflow dataset compared to the SPC dataset (Figs.~\ref{fig:storm_relative_hodographs},~\ref{fig:MLWboxplot}c). 
SRH in the 1--3 km layer in the GR-S HRRR-Inflow has a TSS$_{max}$=0.21, while the SPC SFCOA dataset has essentially no forecast skill (Table~\ref{table:TSSmax}). The tendency of the low-to-mid-level wind profile to possess substantially more forecast skill than any previous model-based tornado climatology holds true regardless of regionality in the GR-S dataset. SRF13, BWD13, and SRH13 all display much higher values of TSS$_{max}$ in the GR-S HRRR-Inflow compared to SPC SFCOA even when filtering events to only the Great Plains or southeastern regions of the United States (not shown), an important distinction given the potential geographic influences in any dataset \citepalias{coniglio2024sampling}.   

As in \citetalias{coniglio2024sampling}, we also find that forecast skill for ground-relative winds (GRWs) aloft is quite high relative to other parameters (Table~\ref{table:TSSmax}). The GRW6 displays the highest skill of any variable tested (TSS$_{max}$ = 0.692; Table~\ref{table:TSSmax}), and most GRWs above the near-ground layer are also quite skillful (i.e., 500 m to  3 km AGL, TSS$_{GRW500mt3km}$ = 0.689). Strong tornadoes and tornado outbreaks have long been associated with anomalously fast low-level and upper-level jets and large surface pressure gradients \citep{johns1992severe}. The high forecast skill of GRWs (from roughly 1--6 km AGL) may be solely related to that, but also could indicate something poorly understood about environments, or the synoptic setup, that favors significant tornadoes [i.e., precipitation distribution \citep{brooks1994role,warren2017impact}, inflow vertical vorticity production in the surface layer \citep{markowski2024new}, or accelerations of the near-storm wind field in response to neighboring convection \citep{werkema2022multistorm}].

The environmental ingredient combination of MLCAPE, MLCIN, MLLCL, EBWD, and some measure of SRH predictably results in the well-documented skill of the STP in the GR-S HRRR-Inflow dataset. Interestingly, the components of the STP variant with the SRH500 (STP500) favorably combine in the GR-S HRRR-Inflow dataset to yield greater forecast skill than any individual component (Table~\ref{table:TSSmax}). In \citetalias{coffer2019srh500}, although STP500 resulted in 8\% more correctly predicted events and 18\% less false alarms than the STP, SRH500 was still more skillful alone than any STP variant. Combining favorable indices into a single parameter more skillful than the individual components is part of the fundamental philosophy of parameter based forecasting \citep[which is not without its own flaws;][]{doswell2006indices} and in the GR-S HRRR-Inflow dataset, this appears to hold true. This may be due to SPC dataset predominately consisting of significantly severe nontornadic supercells or from the regional and seasonal tendencies of the SPC dataset noted in Section~\ref{GRSvsSPC}a; however, the other four ingredients (MLCAPE, MLCIN, MLLCL, and EBWD) in the STP also all have substantially higher TSS$_{max}$ in the GR-S HRRR-Inflow dataset than in the SPC SFCOA dataset. In all, despite more apparent skill within the low-to-mid-level wind profile in the GR-S HRRR-Inflow, there is still forecast value in replacing ESRH with the SRH500 in the formulation of STP500 (the number of correctly predicted events increases by 12\% and the number of missed events/false alarms decreases by 21\%). 


In summary, when looking at the full GR-S HRRR-Inflow dataset, there are noticeable differences within the low-to-mid-level wind profile between nontornadic, weakly tornadic, and significantly tornadic supercells. This is true in terms of storm-relative flow, bulk wind shear, and storm-relative helicity within 1--3 km AGL (similar to the results shown in Section 3), even when using the Bunkers storm motion estimate. Forecast skill of these three components of the low-to-mid-level wind profile in discriminating between nontornadic and significantly tornadic supercells has not been previously seen in prior model-based tornado climatologies and mirrors differences between nontornadic and tornadic supercells in observed field project climatologies \citepalias{coniglio2020insights}. It appears that \emph{some combination of improved resolution and/or representation of the near-storm environment within the HRRR bridges the gap between studies that use soundings from `less sophisticated' model-based analyses (as in \citetalias{coffer2019srh500}) with those observed soundings aggregated from targeted field projects (as in \citetalias{coniglio2020insights})}. While using the HRRR alone does not contradict \citetalias{coffer2019srh500}'s finding regarding the primacy of SRH500 as a tornado predictor, it does recover more of the signal seen in \citetalias{coniglio2020insights}. Next we ask how much additional signal can be extracted from using actual storm motion vectors instead of an a priori estimate. 

\section{Storm motion influence on storm-relative variables}\label{GRSvsSPC-SM}

One of the primary goals at the outset of this study was to create an independent dataset of supercells using GR-S. The provided, observed storm motions in GR-S could be incorporated into the calculations of storm-relative variables, a feature that was unfortunately not available for the larger SPC SFCOA dataset used in previous studies. In this section, we compare bulk TSS$_{max}$ statistics for storm-relative variables and composite storm-relative hodographs exclusively within the GR-S HRRR-Inflow dataset using observed storm motions versus Bunkers storm motion estimates. 

\begin{figure*}[t]
\centerline{\includegraphics[width=40pc]{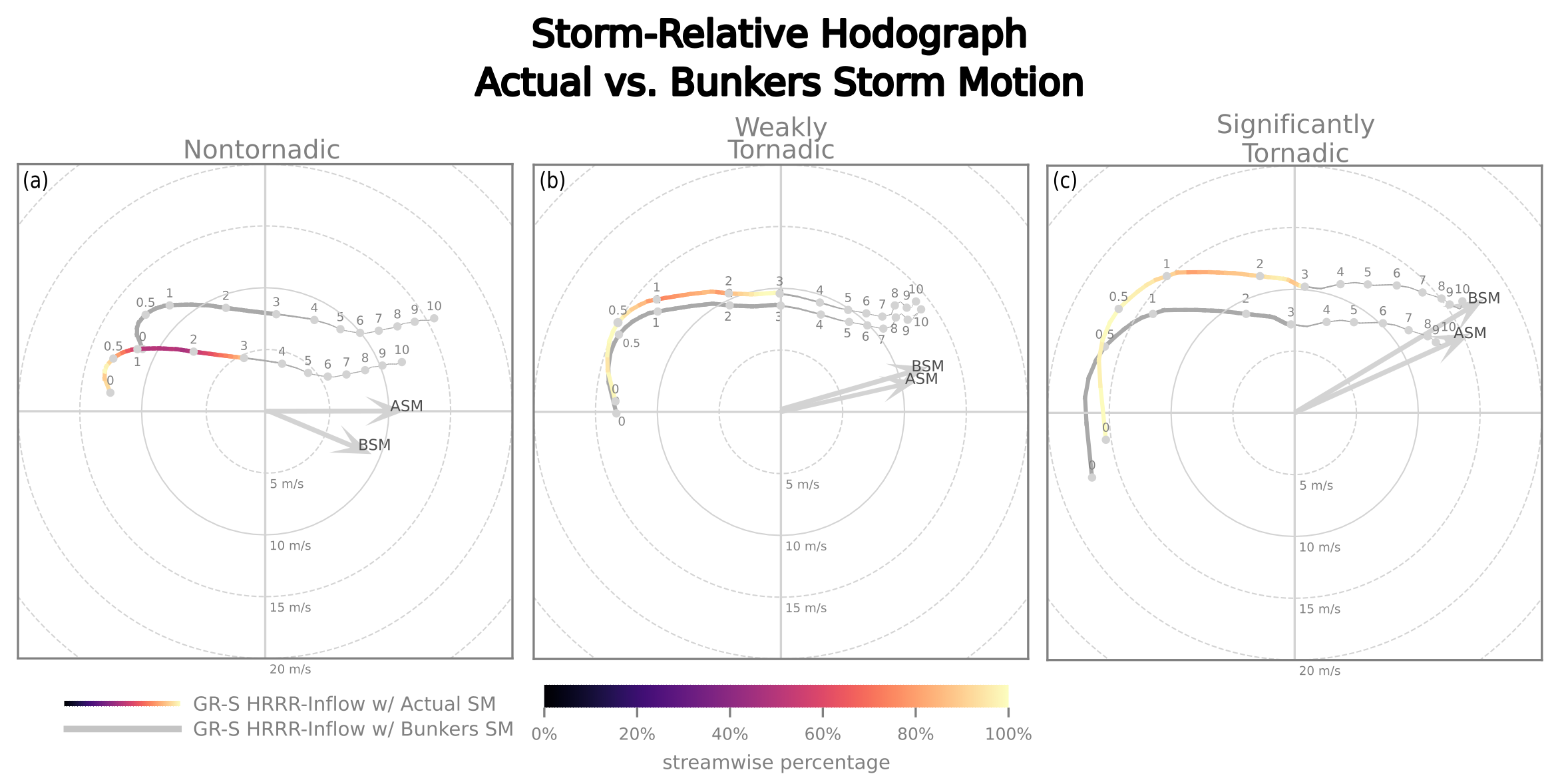}}
\caption{As in Figure~\ref{fig:storm_relative_hodographs}, except all hodographs are from the GR-S HRRR-Inflow dataset and storm motion is calculated from the Bunkers storm motion estimate (BSM; gray) and the actual, observed storm motion (ASM; shaded). The gray BSM hodographs are replicated from the color shaded hodographs in Figure~\ref{fig:storm_relative_hodographs} for comparison. Storm motion vectors for Bunkers and actual storm motions are plotted for visualization purposes.}
\label{fig:storm_relative_hodographs_ASM}
\end{figure*}

On average, in the GR-S HRRR-Inflow dataset, \emph{observed storm motions for tornadic supercells are substantially slower and slightly to the right of Bunkers storm motion estimates, while for nontornadic supercells, the observed motion is slightly faster and substantially to the left} (Table~\ref{table:SMerrors} and shown graphically in Fig.~\ref{fig:storm_relative_hodographs_ASM}). The direction of actual storm motions for tornadic supercells are on average 4$^{\circ}$ to the right of the Bunkers motion estimates, increasing to 6$^{\circ}$ for significantly tornadic supercells, compared to 19$^{\circ}$ to the left of Bunkers motion estimates for the nontornadic subset (Table~\ref{table:SMerrors}). In general, these storm motion errors from GR-S mirror those described by \citet{bunkers2018observations} and \citet{flournoy2021motion}. The Bunkers storm motion estimate for many nontornadic supercells have quite substantial errors in direction compared to the actual storm motion vectors, with 46\% of cases moving at least 20$^{\circ}$ (and 12\% over 40$^{\circ}$) to the left of Bunkers' estimate. Differences in speed between the observed storm motion and Bunkers motion are only $+$0.5 m s$^{-1}$ for the nontornadic subset and $-$0.8m s$^{-1}$ for the weakly tornadic supercells but increase in magntidue to $-$2.1 m s$^{-1}$ for significantly tornadic supercells (Table~\ref{table:SMerrors}). Based on Figure~\ref{fig:SRHexample}, for a given wind profile, these differences in storm motion for tornadic supercells should lead to an enhancement in SRH over deeper layers (with minimal changes in near-ground SRH) with the opposite trends for nontornadic supercells (also discussed by \citetalias{coniglio2020insights}). 

\begin{table*}[]
\caption{Errors in Bunkers storm motion direction (DIR) and speed (SPD) for nontornadic (nontor), weakly tornadic (weaktor), and significantly tornadic supercells (sigtor) compared to the actual storm motion, expressed in mean difference (MD) and mean absolute difference (MAD). For this comparison, difference is synonymous with error.}
\label{table:SMerrors}
\centering
\begin{tabular}{|r|cc|cc|cc|}
\multicolumn{1}{|l|}{} & \multicolumn{2}{c|}{nontor}        & \multicolumn{2}{c|}{weaktor}        & \multicolumn{2}{c|}{sigtor}         \\ \cline{2-7} 
\multicolumn{1}{|l|}{} & \multicolumn{1}{c|}{DIR}  & SPD  & \multicolumn{1}{c|}{DIR}   & SPD  & \multicolumn{1}{c|}{DIR}  & SPD  \\ \hline
MD                     & \multicolumn{1}{c|}{-19.1} & 0.53  & \multicolumn{1}{c|}{1.4}    & -0.75 & \multicolumn{1}{c|}{6.1}   & -2.1   \\ \hline
MAD                    & \multicolumn{1}{c|}{33.3}  & 3.9   & \multicolumn{1}{c|}{19.3}   & 3.4   & \multicolumn{1}{c|}{15.4}  & 3.5    \\ \hline
\end{tabular}
\end{table*}

By using the observed storm motion instead of the Bunkers motion estimates, the forecast skill when comparing nontornadic to significantly tornadic supercells for deeper layers of SRH increases, including SRH3 and ESRH, while the forecast skill for shallower layers decreases, such as SRH500 and SRH1 (Table~\ref{table:TSSmax}). This relative trend of deeper layers gaining skill and shallower layers losing skill is qualitatively similar to \citetalias{coniglio2020insights} and \citetalias{coniglio2024sampling} when using actual storm motions. In \citetalias{coniglio2020insights}, the forecast skill between nontornadic and tornadic supercells increased for SRH3 and ESRH, whereas SRH500 and SRH1 stayed about the same, and SRH100 and SRH250 decreased for the tornadic supercells\footnote{The use of observed soundings in \citetalias{coniglio2020insights} allowed for the calculation of SRH in even shallower layers (i.e., 100 and 250 m AGL) than possible herein where the underlying models used in both the SPC dataset and GR-S HRRR-Inflow only have a handful of grid levels below 500 m AGL.}.

\begin{figure*}[t]
\centerline{\includegraphics[width=30pc]{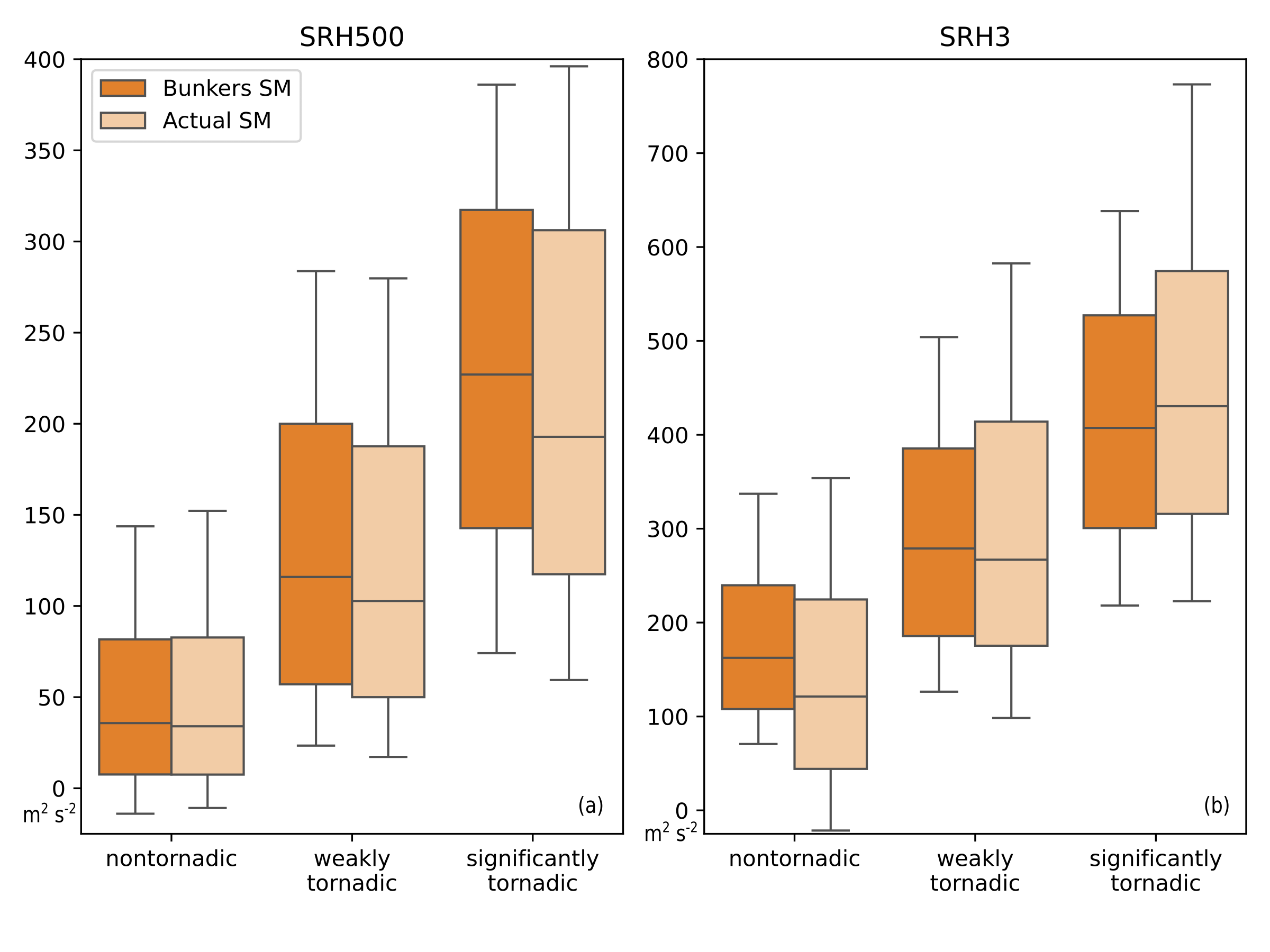}}
\caption{As in Figure~\ref{fig:SRHboxplot}, except for environmental indices derived from the Bunkers storm motion estimate (orange) and observed storm motion (tan) from GR-S HRRR-Inflow dataset.}
\label{fig:SRHboxplot_ASM}
\end{figure*}

In this study, SRH500 is noticeably lower for tornadic storms but essentially unchanged for nontornadic storms when using the actual storm motion (Fig.~\ref{fig:SRHboxplot_ASM}a), leading to a substantial reduction of TSS$_{max}$ for SRH500 (Table~\ref{table:TSSmax}). On the other hand, the distribution of SRH3 is marginally shifted higher for significantly tornadic supercells and noticeably skewed lower for nontornadic supercells (Fig.~\ref{fig:SRHboxplot_ASM}b). Again, the combination of these factors results in a substantial increase in TSS$_{max}$ for SRH3 (Table~\ref{table:TSSmax}). In total, \emph{using observed storm motions, instead of Bunkers' estimates, reverses the trend of progressively shallower layers of SRH having more forecast skill, where now SRH3 has higher TSS$_{max}$ values than SRH500.} Rather interestingly though, as with the composite parameters computed using Bunkers storm motion, STP500 still displays more skill than STP in the GR-S HRRR-Inflow dataset, despite the individual component of ESRH possessing more skill than SRH500 (Table~\ref{table:TSSmax}). Presumably this indicates the other four ingredients in the STP (MLCAPE, MLCIN, MLLCL, EBWD) more positively complement SRH500 than they do ESRH, even when using observed storm motions. Therefore, forecasters may still find value in using STP500 over other variations of STP, despite the increase in forecast skill within the low-to-mid-level wind profile (associated with using the HRRR and observed storm motions). The forecasting ramifications of observed storm motions are discussed at length in Section~\ref{Conclusions}.

\begin{figure*}[t]
\centerline{\includegraphics[width=30pc]{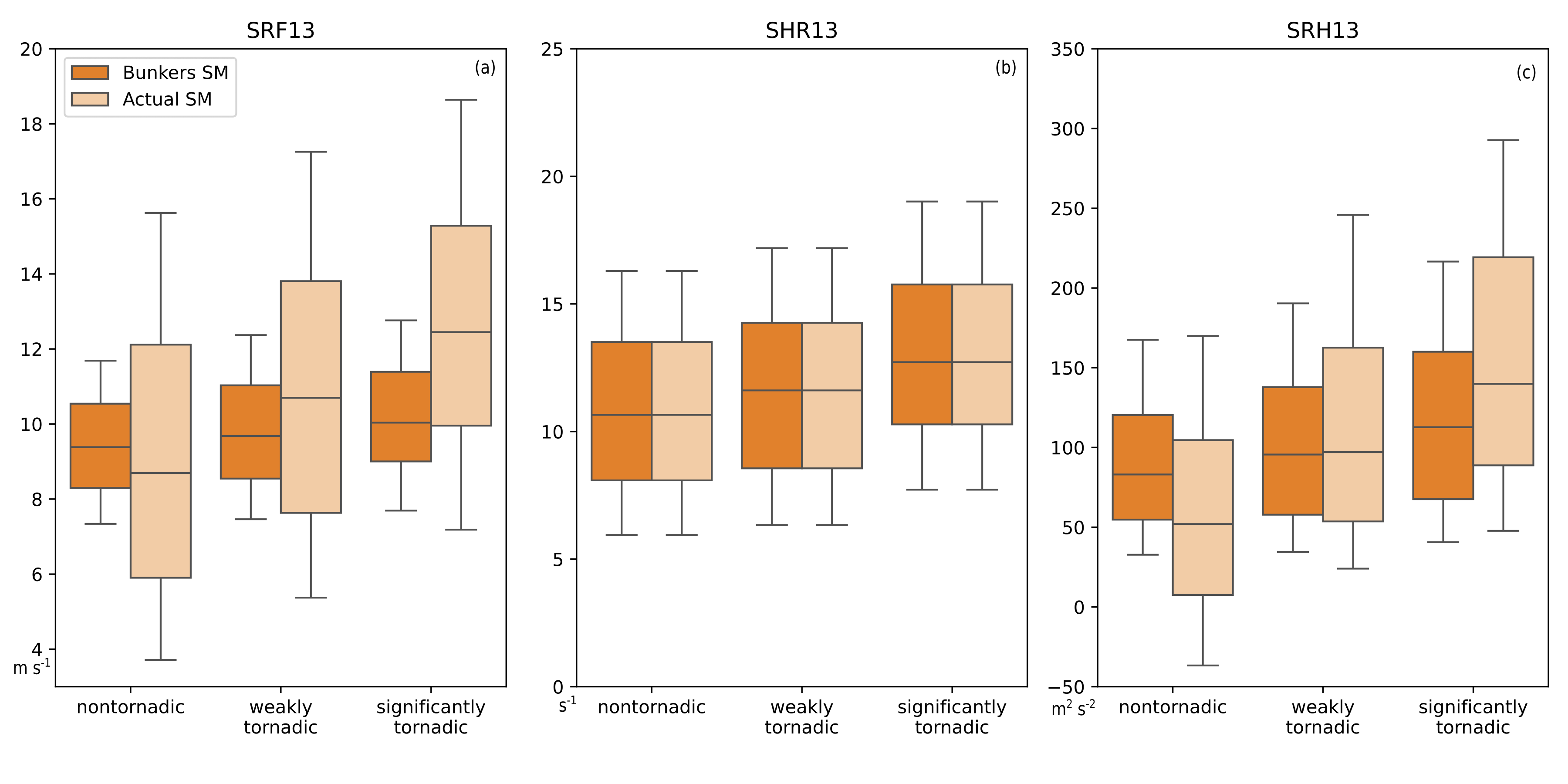}}
\caption{As in Figure~\ref{fig:MLWboxplot}, except for environmental indices derived from the Bunkers storm motion estimate (orange) and observed storm motion (tan) from GR-S HRRR-Inflow dataset.}
\label{fig:MLWboxplot_ASM}
\end{figure*}

We can primarily attribute the reversal of SRH skill versus depth to a further exaggeration of the low-to-mid-level storm-relative wind differences between nontornadic and significantly tornadic supercells. 
Storm-relative winds in the GR-S HRRR-Inflow dataset were already faster in significantly tornadic supercells compared to nontornadic supercells, by 1--2 m s$^{-1}$, using Bunkers storm motion estimates, but this difference is amplified further when using the actual storm motion. With the actual storm motion, SRF13 is slower on average for nontornadic storms by approximately 1 m s$^{-1}$, while in contrast, SRF13 is on average 1 m s$^{-1}$ faster for tornadic storms, further increasing to more than 2-3 m s$^{-1}$ for significantly tornadic supercells (Fig.~\ref{fig:MLWboxplot_ASM}a). Additionally, although the bulk shear in this layer is unchanged (Fig.~\ref{fig:MLWboxplot_ASM}b) as shear is not a storm-relative variable, the changes in storm motion lead to a substantial reduction in the `streamwiseness' of the horizontal vorticity (i.e., the vorticity associated with the vertical wind shear being oriented in streamwise direction) for nontornadic supercells by $\sim$40\% but an inflation in the `streamwiseness' for significantly tornadic supercells (by $\sim$20\% on average; cf. color shaded GR-S HRRR-Inflow hodographs in Figs.~\ref{fig:storm_relative_hodographs},\ref{fig:storm_relative_hodographs_ASM}). The culmination of changes to the storm-relative flow and streamwise percentage results in an even wider gap in the distributions of SRH13 between nontornadic and significantly tornadic supercells for actual storm motions (Fig.~\ref{fig:MLWboxplot_ASM}c). Specifically, TSS$_{max}$ for SRH13 (which was negative in the SPC dataset) increases in the GR-S HRRR-Inflow dataset, with actual storm motions, to 0.45 (Table~\ref{table:TSSmax}). As conceptualized by \citetalias{coniglio2020insights} in the hodograph shown in Figure~\ref{fig:SRHexample}, this explains why deeper layers of SRH (such as SRH3) now display more apparent forecast skill than shallower layers (such as SRH500; Table~\ref{table:TSSmax}). Thus, \emph{we conclude that using actual storm motions from GR-S, combined with near-storm environments from the HRRR, together replicates the full signal seen in \citetalias{coniglio2020insights}, where deeper layers of SRH have higher forecast skill than shallower layers}. The use of the Bunkers storm motion estimate, and documented biases associated with this method, is probably the main factor in why many past tornado climatologies using model-based environmental analyses have failed to find significant differences in the storm-relative wind profile. 

\begin{table*}[]
\caption{Maximum TSS (TSS$_{max}$) values for SRH500 and SRH3 at their respective optimal thresholds (VAR$_{max}$) with the actual storm motion computed over various averaging times ($\Delta$t; minutes) discriminating between significant tornadic supercells (EF2+) and nontornadic supercells within the GridRad-Severe HRRR-Inflow dataset.}
\centering
\label{table:TSSmax-SM}
\begin{tabular}{lcccccc}
\multicolumn{1}{l|}{}          & \multicolumn{6}{c|}{TSS$_{max}$ over varying $\Delta$t (min)}                                                                                 \\ \cline{2-7} 
\multicolumn{1}{l|}{Variables} & 90                   & 60                   & 30                   & 15                   & 5                    & \multicolumn{1}{c|}{+5}    \\ \hline
\multicolumn{1}{c|}{SRH500}    & 0.668                & 0.656                & 0.619                & 0.603                & 0.564                & \multicolumn{1}{c|}{0.502} \\
\multicolumn{1}{c|}{SRH3}      & 0.501                & 0.576                & 0.674                & 0.680                & 0.694                & \multicolumn{1}{c|}{0.689} \\
                               & \multicolumn{1}{l}{} & \multicolumn{1}{l}{} & \multicolumn{1}{l}{} & \multicolumn{1}{l}{} & \multicolumn{1}{l}{} & \multicolumn{1}{l}{}      
\end{tabular}
\end{table*}

In this analysis so far, we have used the storm motion estimate provided along each GR-S storm track, which is a moving average of the prior 30 minutes storm motion based on objective echo top tracking\footnote{Although the time scale of a 30 minute weighted average is similar, this differs from \citetalias{coniglio2020insights}, \citetalias{coniglio2022mesoanalysis}, and \citetalias{coniglio2024sampling} where storm motion was calculated via subjectively tracked updraft locations ``using the location of the inflow notch adjacent to a hook echo on base scan reflectivity, weak echo regions aloft, and/or mesocyclone on radial velocity observations as guidance''.}. 
Notably, \citetalias{coniglio2020insights} showed ``temporary rightward-deviant motions that often precede tornadoes'' for many of the cases (although some cases, such as in \citet{wilson2023environmental}, persisted strongly to the right of Bunkers motion for long periods of time). Short term deviations clearly have implications for interpreting storm-relative variables. In the current study, we also computed storm motions averages over 90, 60, 30, 15, 5, and +5 minute intervals from the raw storm latitude and longitude position provided in GR-S (where +5 minute refers to the five minutes after the event time). In terms of TSS$_{max}$ (Table~\ref{table:TSSmax-SM}), the forecast skill of shallower layers of SRH, such as SRH500, is maximized at the longer averaging times of 60 and 90 minutes, where 90 minutes was most similar to the Bunkers storm motion (SRH500$_{90MIN}$ TSS$_{max}$ = 0.668). Over shorter time periods, the apparent forecast skill of SRH500 decreases, especially between 30 and 15 minutes (SRH500$_{15MIN}$ TSS$_{max}$ = 0.603). The opposite trend is present for deeper layers of SRH. The apparent forecast skill of SRH3 increases with shorter averaging times, with 5 min and +5 min displaying the highest apparent forecast skill (SRH3$_{5MIN}$ TSS$_{max}$ = 0.694). We therefore conclude, similarly to \citetalias{coniglio2020insights}, that \emph{a rightward deviation close to the time of tornadogenesis leads to storm-relative quantities over deeper layers having more forecast skill than shallower layers}. 

\section{Synthesis}\label{Conclusions}

The aim of this article was to use a novel approach to clarify the importance of the storm-relative wind profile in forecasting the likelihood a supercell produces a tornado. Our understanding of the near-storm environmental influence on the internal dynamics of supercell thunderstorms has improved in recent decades. However, the extent which the low-to-mid-level storm-relative wind profile and SRH over various layers has more or less practical forecast skill has become muddled in recent years. 
To address discrepancies in prior work of these specific storm-relative kinematic quantities, an independent, objectively identified dataset of supercells, with observed storm motions, was generated from GridRad-Severe. These supercell events were paired with a high resolution, convection allowing model analysis from the HRRR to represent the complexities of the near-storm inflow environment. The current work supports the following conclusions related to the stated research questions:

\begin{enumerate}

   \item Is the representation of the near-storm environment in HRRR analyses more similar to observations from field projects or to tornado climatologies with traditionally used, coarser model-based analyses?

   \begin{itemize}

     \item The representation of the low-to-mid-level storm relative wind profile using the HRRR appears more similar to observational field project datasets than in coarser model-based analyses over the same set of tornadic supercell events. For most other common convective indices, the HRRR appears to have similar traits and biases as the traditionally used analyses.

   \end{itemize}

   \item If the HRRR analysis is indeed more like previous observational datasets, does the representation of the low-to-mid-level wind profile clarify prior discrepancies of whether shallower or deeper layers of SRH have more forecast skill?

   \begin{itemize}

     \item Within the GridRad-Severe supercell dataset with environments from the HRRR, there are substantive increases in storm-relative flow, vertical wind shear, and storm-relative helicity within the 1--3 km AGL layer for significantly tornadic supercells compared to nontornadic supercells. This yields meaningful forecast skill for these variables, unlike in previous studies using model-based analyses. 

    \item The stronger low-to-mid-level storm-relative flow and vertical wind shear in significantly tornadic supercell environments boosts the forecast skill of deeper layers of SRH in the GridRad-Severe supercell dataset, although SRH500 is still more skillful overall than SRH3. Thus, using a higher resolution analysis alone, at least partly, recovers some of the signal of deeper layers of SRH having more forecast skill from observational field project datasets.

   \end{itemize}


 \item How much do observed storm motions (compared to Bunkers storm motion estimates) alter the forecast skill of storm-relative variables within a large, climatologically representative storm sample?

   \begin{itemize}

     \item Observed storm motions in the GridRad-Severe dataset are slower and to the  right of Bunkers storm motion estimates for tornadic supercells (especially near the time of tornadogenesis), while for nontornadic supercells, the observed storm motion is faster and to the left. 
     \item The differences when using observed storm motion further increase the skill of deeper layers of SRH compared to near-ground layers, ultimately causing deeper layers to have more skill than shallower layers. This seemingly explains the remaining signal of deeper layers of SRH having more forecast skill from observational field project datasets. 

   \end{itemize}

\end{enumerate}

This study represents the culmination of a series of near-storm environment studies by many of the same authors, ranging from \citet{parker2014composite}, \citet{wade2018comparison}, \citetalias{coffer2019srh500}, \citet{coffer2020era5}, \citetalias{coniglio2020insights}, \citetalias{coniglio2022mesoanalysis}, to \citetalias{coniglio2024sampling}. 
How does this all fit together, what have we learned, and what still needs to be clarified? 
Firstly, many uncertainties of the various supercell datasets have been fleshed out. No individual dataset is perfect, and users should be aware of implicit biases in the construction of their dataset. Near-storm observations from this series of studies indicate tornadic storms induce kinematic modifications within the environment that can prove valuable in discriminating themselves from their nontornadic counterparts. In the present study, the HRRR appears to partially capture this effect, where coarser, regional models did not. This may be due to higher resolution, better data assimilation, the inflow averaging technique, or something else about the representation in the HRRR. However, it is not certain what is a ``correct'' background environment, especially if the degree of storm modification is not known ahead of time. 

While this study clarifies certain aspects related to the forecast skill of storm-relative winds and SRH, it also raises questions related to the physical importance of the low-to-mid-level storm-relative winds to internal supercell dynamics. The influence of the near-ground wind profile on the probability of tornadogenesis has a clear, physical explanation related to the intensity of the low-level mesocyclone and the associated dynamic lifting response below cloud base. A similar physical reasoning for the low-to-mid-level winds is still yet to be robustly studied. Is the apparent forecast skill of the low-to-mid-level wind profile a primary driver of a storm's ability to produce a tornado or is it a secondary byproduct? One possibility is that the stronger storm-relative flow, larger shear, and greater SRH in the low-to-mid levels \emph{directly} increase the likelihood of supercell tornadogenesis through effects on the distribution of precipitation or other effects related to the strength of the low-to-mid-level updraft/mesocyclone characteristics. An alternative possibility is that tornadic storms deviate farther to the right prior to tornadogenesis because of an intense low-level mesocyclone (due to large near-ground SRH), which yields the \emph{symptom} of stronger low-to-mid-level storm-relative winds. 
Are supercells' deviant motions a cause or an effect of their propensity to produce a tornado? That is an overarching question this study leaves open to future work. We encourage process-based research on the physical relationship between the low-to-mid-level wind profile and deviant motion in supercells.

Another potential operational ramification of this short time scale deviant motion is that we do not know ahead of time whether a supercell is going to turn right and by how much. On now-casting timescales though, observed storm motions should be used whenever possible. However, without an a priori knowledge of storm motion and access to real-time, high resolution model analyses \citep[such as the Warn-on-Forecast System;][]{heinselman2024warn}, forecasters may still be better off using SRH500 (instead of SRH3), with coarser model analyses and Bunkers storm motion estimates, particularly on the multi-day (i.e., convective outlooks) to hour-long (i.e., severe weather watches and pre-convection initiation mesoscale discussions) timescale prior to severe weather events.  
The increased forecast skill of deeper layers of SRH from deviant motion may require an understanding of the internal storm dynamics that is beyond what we can currently infer prior to convection initiation. Thus, \emph{extracting skill from the low-to-mid-level storm-relative wind profile may prove practically difficult if it requires a highly accurate a priori estimate of storm motion.}

The present study was particularly focused on discrepancies in  the low-to-mid-level wind profile among prior supercell tornado climatologies. In doing so, the benefits of combining GridRad-Severe and the HRRR were demonstrated. Future studies could further exploit these resources to address the environmental influence on hazard likelihood, especially when environmental traits have strong underlying links to the internal dynamics of thunderstorms. These data sources may be especially ripe for more sophisticated, big data techniques to extract new insights about severe weather environments not previously seen in other datasets. We expect that analyses of near-storm environments will continue to provide insights into the prediction and physical processes of tornado formation for some time to come.

\acknowledgments
Authors Coffer and Parker were supported by NSF award AGS-2130936, Homeyer was partially supported by the National Science Foundation under Grant RISE-2019758 (i.e., the NSF AI Institute for Research on Trustworthy AI in Weather, Climate, and Coastal Oceanography [AI2ES]). We thank Dr. Walker Ashley for serving as editor for Weather and Forecasting and Dr. Robert Warren, as well as two anonymous reviewers, for their helpful suggestions. The first author thanks Amanda Murphy and Kiley Allen for answering questions regarding GR-S. Zachary Chalmers is acknowledged for his insights regarding the HRRR, including the representation of near-storm environments across versions and data acquisition. The authors thank John Hart, Ryan Jewell, Bryan Smith, Rich Thompson, and others from the NOAA/NWS Storm Prediction Center for their work on the underlying software used to generate the SPC SFCOA proximity sounding dataset, tireless effort in assigning convective modes to each report, and their generosity for sharing the data with the community. A portion of this work used code provided by Brian Blaylock’s Herbie python package (https://doi.org/10.5281/zenodo.4567540). Many figures and analyses were created using open-sourced Python packages such as Numpy (Harris et al. 2020), Matplotlib (Hunter 2007), Pandas (McKinney 2010), Xarray (Hoyer and Hamman 2017), Metpy (May et al. 2017), and Jupyter (Kluyver et al. 2016). The composite storm-relative hodographs were made with code adapted from Sam Brandt (https://github.com/SamBrandtMeteo/Storm-Relative-Hodograph-Plotter). The composite Skew$T$ log--$p$ diagrams were created using the SounderPy package (Gillet 2024).

\datastatement
The GridRad-Severe dataset can be found at the NCAR Research Data Archive \citep{gridradsevereCISL}. The High Resolution Rapid Refresh is being continuously archived on various cloud platforms, including Amazon Web Server (https://registry.opendata.aws/noaa-hrrr-pds/). The SPC SFCOA dataset are archived at the Storm Prediction Center (SPC) and available by contacting the Science Support Branch at the SPC. 

\bibliographystyle{ametsocV6}
\bibliography{references}

\begin{thebibliography}{87}
\providecommand{\natexlab}[1]{#1}
\providecommand{\url}[1]{\texttt{#1}}
\renewcommand{\UrlFont}{\rmfamily}
\providecommand{\urlprefix}{URL }
\expandafter\ifx\csname urlstyle\endcsname\relax
  \providecommand{\doi}[1]{https://doi.org/\discretionary{}{}{}#1}\else
  \providecommand{\doi}{https://doi.org/\discretionary{}{}{}\begingroup \urlstyle{rm}\Url}\fi
\providecommand{\eprint}[2][]{\url{#2}}

\bibitem[{Anderson-Frey et~al.(2016)Anderson-Frey, Richardson, Dean, Thompson,, and Smith}]{anderson2016investigation}
Anderson-Frey, A.~K., Y.~P. Richardson, A.~R. Dean, R.~L. Thompson, and B.~T. Smith, 2016: Investigation of near-storm environments for tornado events and warnings. \textit{Wea.\ Forecasting}, \textbf{31~(6)}, 1771--1790.

\bibitem[{Beebe(1958)}]{beebe1958tornado}
Beebe, R.~G., 1958: Tornado proximity soundings. \textit{Bull.\ Amer.\ Meteor.\ Soc.}, \textbf{39~(4)}, 195--201.

\bibitem[{Benjamin et~al.(2004)}]{benjamin2004ruc}
Benjamin, S.~G., and Coauthors, 2004: An hourly assimilation--forecast cycle: The {RUC}. \textit{Mon.\ Wea.\ Rev.}, \textbf{132~(2)}, 495--518.

\bibitem[{Benjamin et~al.(2016)}]{benjamin2016rap}
Benjamin, S.~G., and Coauthors, 2016: A {N}orth {A}merican hourly assimilation and model forecast cycle: {T}he {R}apid {R}efresh. \textit{Mon.\ Wea.\ Rev.}, \textbf{144~(4)}, 1669--1694.

\bibitem[{Blaylock(2022)}]{blaylock2022herbie}
Blaylock, B.~K., 2022: Herbie: Retrieve numerical weather prediction model data (version 2022.9.0). \urlprefix\url{https://doi.org/10.5281/zenodo.4567540}.

\bibitem[{Blaylock et~al.(2017)Blaylock, Horel,, and Liston}]{blaylock2017hrrr}
Blaylock, B.~K., J.~D. Horel, and S.~T. Liston, 2017: Cloud archiving and data mining of {H}igh-{R}esolution {R}apid {R}efresh forecast model output. \textit{Computers and {G}eosciences}, \textbf{109~(Supplement C)}, 43 -- 50.

\bibitem[{Bothwell et~al.(2002)Bothwell, Hart,, and Thompson}]{bothwell2002integrated}
Bothwell, P., J.~Hart, and R.~Thompson, 2002: An integrated three-dimensional objective analysis scheme in use at the {S}torm {P}rediction {C}enter. \textit{21st Conf. on Severe Local Storms}, Amer. Meteor. Soc., San Antonio, TX.

\bibitem[{Brooks et~al.(1994{\natexlab{a}})Brooks, Doswell,, and Wilhelmson}]{brooks1994role}
Brooks, H.~E., C.~A. Doswell, and R.~B. Wilhelmson, 1994{\natexlab{a}}: The role of midtropospheric winds in the evolution and maintenance of low-level mesocyclones. \textit{Mon.\ Wea.\ Rev.}, \textbf{122~(1)}, 126--136.

\bibitem[{Brooks et~al.(1994{\natexlab{b}})Brooks, Doswell~III,, and Cooper}]{brooks1994environments}
Brooks, H.~E., C.~A. Doswell~III, and J.~Cooper, 1994{\natexlab{b}}: On the environments of tornadic and nontornadic mesocyclones. \textit{Wea.\ Forecasting}, \textbf{9~(4)}, 606--618.

\bibitem[{Bunkers(2018)}]{bunkers2018observations}
Bunkers, M.~J., 2018: Observations of right-moving supercell motion forecast errors. \textit{Wea.\ Forecasting}, \textbf{33~(1)}, 145--159.

\bibitem[{Bunkers et~al.(2000)Bunkers, Klimowski, Zeitler, Thompson,, and Weisman}]{bunkers2000predicting}
Bunkers, M.~J., B.~A. Klimowski, J.~W. Zeitler, R.~L. Thompson, and M.~L. Weisman, 2000: Predicting supercell motion using a new hodograph technique. \textit{Wea.\ Forecasting}, \textbf{15~(1)}, 61--79.

\bibitem[{Burlingame et~al.(2017)Burlingame, Evans,, and Roebber}]{burlingame2017pbl}
Burlingame, B.~M., C.~Evans, and P.~J. Roebber, 2017: The influence of {PBL} parameterization on the practical predictability of convection initiation during the {M}esoscale {P}redictability {EX}periment ({MPEX}). \textit{Wea.\ Forecasting}, \textbf{32~(3)}, 1161--1183.

\bibitem[{Clark et~al.(2023)}]{clark2023sfe}
Clark, A.~J., and Coauthors, 2023: The first hybrid {NOAA} {H}azardous {W}eather {T}estbed {S}pring {F}orecasting {E}xperiment for advancing severe weather prediction. \textit{Bull.\ Amer.\ Meteor.\ Soc.}, \textbf{104~(12)}, E2305--E2307.

\bibitem[{Coffer and Parker(2017)Coffer, and Parker}]{coffer2017simulated}
Coffer, B.~E., and M.~D. Parker, 2017: Simulated supercells in nontornadic and tornadic {VORTEX2} environments. \textit{Mon.\ Wea.\ Rev.}, \textbf{145~(1)}, 149--180.

\bibitem[{Coffer and Parker(2018)Coffer, and Parker}]{coffer2018tipping}
Coffer, B.~E., and M.~D. Parker, 2018: Is there a “tipping point” between simulated nontornadic and tornadic supercells in {VORTEX2} environments? \textit{Mon.\ Wea.\ Rev.}, \textbf{146~(8)}, 2667--2693.

\bibitem[{Coffer et~al.(2017)Coffer, Parker, Dahl, Wicker,, and Clark}]{coffer2017volatility}
Coffer, B.~E., M.~D. Parker, J.~M. Dahl, L.~J. Wicker, and A.~J. Clark, 2017: Volatility of tornadogenesis: An ensemble of simulated nontornadic and tornadic supercells in {VORTEX2} environments. \textit{Mon.\ Wea.\ Rev.}, \textbf{145~(11)}.

\bibitem[{Coffer et~al.(2023)Coffer, Parker, Peters,, and Wade}]{coffer2023LLM}
Coffer, B.~E., M.~D. Parker, J.~M. Peters, and A.~R. Wade, 2023: Supercell low-level mesocyclones: {O}rigins of inflow and vorticity. \textit{Mon.\ Wea.\ Rev.}, \textbf{151~(9)}, 2205--2232.

\bibitem[{Coffer et~al.(2019)Coffer, Parker, Thompson, Smith,, and Jewell}]{coffer2019srh500}
Coffer, B.~E., M.~D. Parker, R.~L. Thompson, B.~T. Smith, and R.~E. Jewell, 2019: Using near-ground storm relative helicity in supercell tornado forecasting. \textit{Wea.\ Forecasting}, \textbf{34~(5)}, 1417--1435.

\bibitem[{Coffer et~al.(2020)Coffer, Taszarek,, and Parker}]{coffer2020era5}
Coffer, B.~E., M.~Taszarek, and M.~D. Parker, 2020: Near-ground wind profiles of tornadic and nontornadic environments in the {U}nited {S}tates and {E}urope from {ERA5} reanalyses. \textit{Wea.\ Forecasting}, \textbf{35~(6)}, 2621--2638.

\bibitem[{Coniglio(2012)}]{coniglio2012verification}
Coniglio, M.~C., 2012: Verification of {RUC} 0-1-h forecasts and {SPC} mesoscale analyses using {VORTEX2} soundings. \textit{Wea.\ Forecasting}, \textbf{27~(3)}, 667--683.

\bibitem[{Coniglio and Jewell(2022)Coniglio, and Jewell}]{coniglio2022mesoanalysis}
Coniglio, M.~C., and R.~E. Jewell, 2022: {SPC} mesoscale analysis compared to field-project soundings: {I}mplications for supercell environment studies. \textit{Mon.\ Wea.\ Rev.}, \textbf{150~(3)}, 567--588.

\bibitem[{Coniglio and Parker(2020)Coniglio, and Parker}]{coniglio2020insights}
Coniglio, M.~C., and M.~D. Parker, 2020: Insights into supercells and their environments from three decades of targeted radiosonde observations. \textit{Monthly Weather Review}, \textbf{148~(12)}, 4893--4915.

\bibitem[{Coniglio and Thompson(2024)Coniglio, and Thompson}]{coniglio2024sampling}
Coniglio, M.~C., and R.~L. Thompson, 2024: Impacts of sampling and storm-motion estimates on {RUC/RAP}-based discriminations of nontornadic and tornadic supercell environments. \textit{Wea.\ Forecasting}, \textbf{39~(10)}, 1417--1434.

\bibitem[{Cooney et~al.(2018)Cooney, Bowman, Homeyer,, and Fenske}]{cooney2018ten}
Cooney, J.~W., K.~P. Bowman, C.~R. Homeyer, and T.~M. Fenske, 2018: Ten year analysis of tropopause-overshooting convection using {G}rid{R}ad data. \textit{Journal of Geophysical Research: Atmospheres}, \textbf{123~(1)}, 329--343.

\bibitem[{Craven et~al.(2004)Craven, Brooks,, and Hart}]{craven2004baseline}
Craven, J.~P., H.~E. Brooks, and J.~A. Hart, 2004: Baseline climatology of sounding derived parameters associated with deep, moist convection. \textit{Natl.\ Weather Dig.}, \textbf{28~(1)}, 13--24.

\bibitem[{Davenport and Parker(2015)Davenport, and Parker}]{davenport2015impact}
Davenport, C.~E., and M.~D. Parker, 2015: Impact of environmental heterogeneity on the dynamics of a dissipating supercell thunderstorm. \textit{Mon.\ Wea.\ Rev.}, \textbf{143~(10)}, 4244--4277.

\bibitem[{Davies-Jones(2015)}]{davies2015review}
Davies-Jones, R., 2015: A review of supercell and tornado dynamics. \textit{Atmos.\ Res.}, \textbf{158--159}, 274--291.

\bibitem[{Davis and Parker(2014)Davis, and Parker}]{davis2014radar}
Davis, J.~M., and M.~D. Parker, 2014: Radar climatology of tornadic and nontornadic vortices in high-shear, low-cape environments in the mid-atlantic and southeastern united states. \textit{Wea.\ Forecasting}, \textbf{29~(4)}, 828--853.

\bibitem[{Doswell and Schultz(2006)Doswell, and Schultz}]{doswell2006indices}
Doswell, C.~A., III, and D.~M. Schultz, 2006: On the use of indices and parameters in forecasting severe storms. \textit{Electronic J. Severe Storms Meteor.}, \textbf{1~(3)}, 1559--5404.

\bibitem[{Dowell et~al.(2022)}]{dowell2022hrrr}
Dowell, D.~C., and Coauthors, 2022: The {H}igh-{R}esolution {R}apid {R}efresh ({HRRR}): An hourly updating convection-allowing forecast model. {P}art {I}: {M}otivation and system description. \textit{Wea.\ Forecasting}, \textbf{37~(8)}, 1371--1395.

\bibitem[{Fawbush and Miller(1952)Fawbush, and Miller}]{fawbush1952mean}
Fawbush, W., and R.~Miller, 1952: A mean sounding representative of the tornadic airmass environment. \textit{Bull.\ Amer.\ Meteor.\ Soc.}, \textbf{33~(7)}, 303--307.

\bibitem[{Fischer et~al.(2024)Fischer, Dahl, Coffer, Houser, Markowski, Parker, Weiss,, and Schueth}]{fischer2024progress}
Fischer, J., J.~M. Dahl, B.~E. Coffer, J.~L. Houser, P.~M. Markowski, M.~D. Parker, C.~C. Weiss, and A.~Schueth, 2024: Supercell tornadogenesis: Recent progress in our state of understanding. \textit{Bulletin of the American Meteorological Society}, \textbf{105~(7)}, E1084 -- E1097.

\bibitem[{Flournoy et~al.(2021)Flournoy, Coniglio,, and Rasmussen}]{flournoy2021motion}
Flournoy, M.~D., M.~C. Coniglio, and E.~N. Rasmussen, 2021: Examining relationships between environmental conditions and supercell motion in time. \textit{Wea.\ Forecasting}, \textbf{36~(3)}, 737--755.

\bibitem[{Gensini and Brooks(2018)Gensini, and Brooks}]{gensini2018spatial}
Gensini, V.~A., and H.~E. Brooks, 2018: Spatial trends in {U}nited {S}tates tornado frequency. \textit{NPJ Climate and Atmospheric Science}, \textbf{1~(1)}, 38.

\bibitem[{Goldacker and Parker(2021)Goldacker, and Parker}]{goldacker2021updraft}
Goldacker, N.~A., and M.~D. Parker, 2021: Low-level updraft intensification in response to environmental wind profiles. \textit{J.\ Atmos.\ Sci.}, \textbf{78~(9)}, 2763--2781.

\bibitem[{Goldacker and Parker(2023)Goldacker, and Parker}]{goldacker2023srh}
Goldacker, N.~A., and M.~D. Parker, 2023: Assessing the comparative effects of storm-relative helicity components within right-moving supercell environments. \textit{J.\ Atmos.\ Sci.}, \textbf{80~(12)}, 2805--2822.

\bibitem[{Gray and Frame(2021)Gray, and Frame}]{gray2021midlevel}
Gray, K., and J.~Frame, 2021: The impact of midlevel shear orientation on the longevity of and downdraft location and tornado-like vortex formation within simulated supercells. \textit{Mon.\ Wea.\ Rev.}, \textbf{149~(11)}, 3739--3759.

\bibitem[{{Grid{R}ad-{S}evere}(2021)}]{gridradsevereCISL}
{Grid{R}ad-{S}evere}, 2021: Grid{R}ad-{S}evere - {T}hree-dimensional gridded {NEXRAD WSR-88D} radar data for severe events. Research {D}ata {A}rchive at the {N}ational {C}enter for {A}tmospheric {R}esearch, {C}omputational and {I}nformation {S}ystems {L}aboratory, Boulder {CO}, \urlprefix\url{https://doi.org/10.5065/2B46-1A97}.

\bibitem[{Heinselman et~al.(2024)}]{heinselman2024warn}
Heinselman, P.~L., and Coauthors, 2024: Warn-on-{F}orecast {S}ystem: {F}rom vision to reality. \textit{Wea.\ Forecasting}, \textbf{39~(1)}, 75--95.

\bibitem[{Herman et~al.(2018)Herman, Nielsen,, and Schumacher}]{herman2018probabilistic}
Herman, G.~R., E.~R. Nielsen, and R.~S. Schumacher, 2018: Probabilistic verification of {S}torm {P}rediction {C}enter convective outlooks. \textit{Wea.\ Forecasting}, \textbf{33~(1)}, 161--184.

\bibitem[{Hitchens and Brooks(2014)Hitchens, and Brooks}]{hitchens2014evaluation}
Hitchens, N.~M., and H.~E. Brooks, 2014: Evaluation of the {S}torm {P}rediction {C}enter’s convective outlooks from day 3 through day 1. \textit{Wea.\ Forecasting}, \textbf{29~(5)}, 1134--1142.

\bibitem[{Homeyer and Bowman(2022)Homeyer, and Bowman}]{GridRaddataset}
Homeyer, C.~R., and K.~Bowman, 2022: Algorithm description document for version 4.2 of the three-dimensional gridded {NEXRAD WSR-88D} radar ({G}rid{R}ad) dataset. Tech {D}oc, \urlprefix\url{http://gridrad.org/pdf/GridRad-v4.2-Algorithm-Description.pdf}, 33pp pp.

\bibitem[{Homeyer et~al.(2017)Homeyer, McAuliffe,, and Bedka}]{homeyer2017cirrus}
Homeyer, C.~R., J.~D. McAuliffe, and K.~M. Bedka, 2017: On the development of above-anvil cirrus plumes in extratropical convection. \textit{J.\ Atmos.\ Sci.}, \textbf{74~(5)}, 1617--1633.

\bibitem[{Homeyer et~al.(2020)Homeyer, Sandm{\ae}l, Potvin,, and Murphy}]{homeyer2020distinguishing}
Homeyer, C.~R., T.~N. Sandm{\ae}l, C.~K. Potvin, and A.~M. Murphy, 2020: Distinguishing characteristics of tornadic and nontornadic supercell storms from composite mean analyses of radar observations. \textit{Mon.\ Wea.\ Rev.}, \textbf{148~(12)}, 5015--5040.

\bibitem[{Jahn et~al.(2020)Jahn, Gallo, Broyles, Smith, Jirak,, and Milne}]{jahn2020inflow}
Jahn, D.~E., B.~T. Gallo, C.~Broyles, B.~T. Smith, I.~Jirak, and J.~Milne, 2020: Refining {CAM}-based tornado probability forecasts using storm-inflow and storm-attribute information. \textit{26th Conf. Numerical Weather Prediction, Boston, MA, Amer. Meteor. Soc}, Vol. 2A.4.

\bibitem[{James et~al.(2022)}]{james2022hrrr}
James, E.~P., and Coauthors, 2022: The {H}igh-{R}esolution {R}apid {R}efresh ({HRRR}): an hourly updating convection-allowing forecast model. part {II}: {F}orecast performance. \textit{Wea.\ Forecasting}, \textbf{37~(8)}, 1397--1417.

\bibitem[{Johns and Doswell(1992)Johns, and Doswell}]{johns1992severe}
Johns, R.~H., and C.~A. Doswell, III, 1992: Severe local storms forecasting. \textit{Wea.\ Forecasting}, \textbf{7~(4)}, 588--612.

\bibitem[{Kerr et~al.(2019)Kerr, Stensrud,, and Wang}]{kerr2019diagnosing}
Kerr, C.~A., D.~J. Stensrud, and X.~Wang, 2019: Diagnosing convective dependencies on near-storm environments using ensemble sensitivity analyses. \textit{Mon.\ Wea.\ Rev.}, \textbf{147~(2)}, 495--517.

\bibitem[{Lagerquist et~al.(2020)Lagerquist, McGovern, Homeyer, Gagne~II,, and Smith}]{lagerquist2020deep}
Lagerquist, R., A.~McGovern, C.~R. Homeyer, D.~J. Gagne~II, and T.~Smith, 2020: Deep learning on three-dimensional multiscale data for next-hour tornado prediction. \textit{Mon.\ Wea.\ Rev.}, \textbf{148~(7)}, 2837--2861.

\bibitem[{Laser et~al.(2022)Laser, Coniglio, Skinner,, and Smith}]{laser2022doppler}
Laser, J.~J., M.~C. Coniglio, P.~S. Skinner, and E.~N. Smith, 2022: Doppler lidar and mobile radiosonde observation-based evaluation of {W}arn-on-{F}orecast {S}ystem predicted near-supercell environments during {TORUS} 2019. \textit{Wea.\ Forecasting}, \textbf{37~(10)}, 1783--1804.

\bibitem[{Maddox(1976)}]{maddox1976evaluation}
Maddox, R.~A., 1976: An evaluation of tornado proximity wind and stability data. \textit{Mon.\ Wea.\ Rev.}, \textbf{104~(2)}, 133--142.

\bibitem[{Markowski(2024)}]{markowski2024new}
Markowski, P.~M., 2024: A new pathway for tornadogenesis exposed by numerical simulations of supercells in turbulent environments. \textit{J.\ Atmos.\ Sci.}, \textbf{81~(3)}, 481--518.

\bibitem[{Markowski et~al.(2003)Markowski, Hannon, Frame, Lancaster, Pietrycha, Edwards,, and Thompson}]{markowski2003characteristics}
Markowski, P.~M., C.~Hannon, J.~Frame, E.~Lancaster, A.~Pietrycha, R.~Edwards, and R.~L. Thompson, 2003: Characteristics of vertical wind profiles near supercells obtained from the {R}apid {U}pdate {C}ycle. \textit{Wea.\ Forecasting}, \textbf{18~(6)}, 1262--1272.

\bibitem[{Markowski and Richardson(2009)Markowski, and Richardson}]{markowski2009tornadogenesis}
Markowski, P.~M., and Y.~P. Richardson, 2009: Tornadogenesis: Our current understanding, forecasting considerations, and questions to guide future research. \textit{Atmos.\ Res.}, \textbf{93~(1-3)}, 3--10.

\bibitem[{Markowski and Richardson(2014)Markowski, and Richardson}]{markowski2014influence}
Markowski, P.~M., and Y.~P. Richardson, 2014: The influence of environmental low-level shear and cold pools on tornadogenesis: Insights from idealized simulations. \textit{J.\ Atmos.\ Sci.}, \textbf{71~(1)}, 243--275.

\bibitem[{Markowski et~al.(1998)Markowski, Straka, Rasmussen,, and Blanchard}]{markowski1998variability}
Markowski, P.~M., J.~M. Straka, E.~N. Rasmussen, and D.~O. Blanchard, 1998: Variability of storm-relative helicity during {VORTEX}. \textit{Mon.\ Wea.\ Rev.}, \textbf{126~(11)}, 2959--2971.

\bibitem[{Muehr et~al.(2024)Muehr, Ruppert, Flournoy,, and Peters}]{muehr2024influence}
Muehr, A.~J., J.~H. Ruppert, M.~D. Flournoy, and J.~M. Peters, 2024: The influence of midlevel shear and horizontal rotors on supercell updraft dynamics. \textit{J.\ Atmos.\ Sci.}, \textbf{81~(1)}, 153--176.

\bibitem[{Mulholland et~al.(2024)Mulholland, Nowotarski, Peters, Morrison,, and Nielsen}]{mulholland2024vws}
Mulholland, J.~P., C.~J. Nowotarski, J.~M. Peters, H.~Morrison, and E.~R. Nielsen, 2024: How does vertical wind shear influence updraft characteristics and hydrometeor distributions in supercell thunderstorms? \textit{Mon.\ Wea.\ Rev.}, \textbf{152~(7)}, 1663--1687.

\bibitem[{Murphy et~al.(2023)Murphy, Homeyer,, and Allen}]{murphy2023gridrad}
Murphy, A.~M., C.~R. Homeyer, and K.~Q. Allen, 2023: Development and investigation of {G}rid{R}ad-severe, a multiyear severe event radar dataset. \textit{Mon.\ Wea.\ Rev.}, \textbf{151~(9)}, 2257--2277.

\bibitem[{Nixon and Allen(2022)Nixon, and Allen}]{nixon2022distinguishing}
Nixon, C.~J., and J.~T. Allen, 2022: Distinguishing between hodographs of severe hail and tornadoes. \textit{Wea.\ Forecasting}.

\bibitem[{{NOAA/NCEI}(2022)}]{NOAA2022}
{NOAA/NCEI}, 2022: {NOAA’s storm events database}. \urlprefix\url{https://www.ncdc.noaa.gov/stormevents/}, {National Centers for Environmental Information, accessed December 2019 to January 2022}.

\bibitem[{Nowotarski and Jensen(2013)Nowotarski, and Jensen}]{nowotarski2013classifying}
Nowotarski, C.~J., and A.~A. Jensen, 2013: Classifying proximity soundings with self-organizing maps toward improving supercell and tornado forecasting. \textit{Wea.\ Forecasting}, \textbf{28~(3)}, 783--801.

\bibitem[{Nowotarski et~al.(2014)Nowotarski, Markowski, Richardson,, and Bryan}]{nowotarski2014properties}
Nowotarski, C.~J., P.~M. Markowski, Y.~P. Richardson, and G.~H. Bryan, 2014: Properties of a simulated convective boundary layer in an idealized supercell thunderstorm environment. \textit{Mon.\ Wea.\ Rev.}, \textbf{142~(11)}, 3955--3976.

\bibitem[{Parker(2014)}]{parker2014composite}
Parker, M.~D., 2014: Composite {VORTEX2} supercell environments from near-storm soundings. \textit{Mon.\ Wea.\ Rev.}, \textbf{142~(2)}, 508--529.

\bibitem[{Parker(2017)}]{parker2017backing}
Parker, M.~D., 2017: How much does “backing aloft” actually impact a supercell? \textit{Wea.\ Forecasting}, \textbf{32~(5)}, 1937--1957.

\bibitem[{Peters et~al.(2023)Peters, Coffer, Parker, Nowotarski, Mulholland, Nixon,, and Allen}]{peters2023disentangling}
Peters, J.~M., B.~E. Coffer, M.~D. Parker, C.~J. Nowotarski, J.~P. Mulholland, C.~J. Nixon, and J.~T. Allen, 2023: Disentangling the influences of storm-relative flow and horizontal streamwise vorticity on low-level mesocyclones in supercells. \textit{J.\ Atmos.\ Sci.}, \textbf{80~(1)}, 129--149.

\bibitem[{Peters et~al.(2019)Peters, Nowotarski,, and Morrison}]{peters2020updraft}
Peters, J.~M., C.~J. Nowotarski, and H.~Morrison, 2019: The role of vertical wind shear in modulating maximum supercell updraft velocities. \textit{J.\ Atmos.\ Sci.}, \textbf{76~(10)}, 3169--3189.

\bibitem[{Potvin et~al.(2010)Potvin, Elmore,, and Weiss}]{potvin2010assessing}
Potvin, C.~K., K.~L. Elmore, and S.~J. Weiss, 2010: Assessing the impacts of proximity sounding criteria on the climatology of significant tornado environments. \textit{Wea.\ Forecasting}, \textbf{25~(3)}, 921--930.

\bibitem[{Potvin et~al.(2019)}]{potvin2019systematic}
Potvin, C.~K., and Coauthors, 2019: Systematic comparison of convection-allowing models during the 2017 {NOAA} {HWT} spring forecasting experiment. \textit{Wea.\ Forecasting}, \textbf{34~(5)}, 1395--1416.

\bibitem[{Potvin et~al.(2020)}]{potvin2020assessing}
Potvin, C.~K., and Coauthors, 2020: Assessing systematic impacts of {PBL} schemes on storm evolution in the {NOAA} {W}arn-on-{F}orecast {S}ystem. \textit{Mon.\ Wea.\ Rev.}, \textbf{148~(6)}, 2567--2590.

\bibitem[{Rasmussen(2003)}]{rasmussen2003refined}
Rasmussen, E.~N., 2003: Refined supercell and tornado forecast parameters. \textit{Wea.\ Forecasting}, \textbf{18~(3)}, 530--535.

\bibitem[{Rasmussen and Blanchard(1998)Rasmussen, and Blanchard}]{rasmussen1998baseline}
Rasmussen, E.~N., and D.~O. Blanchard, 1998: A baseline climatology of sounding-derived supercell and tornado forecast parameters. \textit{Wea.\ Forecasting}, \textbf{13~(4)}, 1148--1164.

\bibitem[{Roberts et~al.(2022)Roberts, Gallo, Jirak,, and Clark}]{roberts2022href}
Roberts, B., B.~T. Gallo, I.~L. Jirak, and A.~J. Clark, 2022: The high resolution ensemble forecast (href) system: Applications and performance for forecasting convective storms. \textit{Authorea Preprints}.

\bibitem[{Sandmael(2017)}]{sandmael2017evaluation}
Sandmael, T., 2017: An evaluation of radar-and satellite-data based products to discriminate between tornadic and non-tornadic stormss. Ph.D. thesis, Univ. of Oklahoma.

\bibitem[{Smith et~al.(2012)Smith, Thompson, Grams, Broyles,, and Brooks}]{smith2012modes}
Smith, B.~T., R.~L. Thompson, J.~S. Grams, C.~Broyles, and H.~E. Brooks, 2012: Convective modes for significant severe thunderstorms in the contiguous {U}nited {S}tates. {P}art {I}: Storm classification and climatology. \textit{Wea.\ Forecasting}, \textbf{27~(5)}, 1114--1135.

\bibitem[{Thompson(1998)}]{thompson1998eta}
Thompson, R.~L., 1998: Eta model storm-relative winds associated with tornadic and nontornadic supercells. \textit{Wea.\ Forecasting}, \textbf{13~(1)}, 125--137.

\bibitem[{Thompson et~al.(2003)Thompson, Edwards, Hart, Elmore,, and Markowski}]{thompson2003close}
Thompson, R.~L., R.~Edwards, J.~A. Hart, K.~L. Elmore, and P.~Markowski, 2003: Close proximity soundings within supercell environments obtained from the {R}apid {U}pdate {C}ycle. \textit{Wea.\ Forecasting}, \textbf{18~(6)}, 1243--1261.

\bibitem[{Thompson et~al.(2007)Thompson, Mead,, and Edwards}]{thompson2007effective}
Thompson, R.~L., C.~M. Mead, and R.~Edwards, 2007: Effective storm-relative helicity and bulk shear in supercell thunderstorm environments. \textit{Wea.\ Forecasting}, \textbf{22~(1)}, 102--115.

\bibitem[{Thompson et~al.(2012)Thompson, Smith, Grams, Dean,, and Broyles}]{thompson2012convective}
Thompson, R.~L., B.~T. Smith, J.~S. Grams, A.~R. Dean, and C.~Broyles, 2012: Convective modes for significant severe thunderstorms in the contiguous {U}nited {S}tates. {P}art {II}: {S}upercell and {QLCS} tornado environments. \textit{Wea.\ Forecasting}, \textbf{27~(5)}, 1136--1154.

\bibitem[{Togstad et~al.(2011)Togstad, Davies, Corfidi, Bright,, and Dean}]{togstad2011conditional}
Togstad, W.~E., J.~M. Davies, S.~J. Corfidi, D.~R. Bright, and A.~R. Dean, 2011: Conditional probability estimation for significant tornadoes based on {R}apid {U}pdate {C}ycle ({RUC}) profiles. \textit{Wea.\ Forecasting}, \textbf{26~(5)}, 729--743.

\bibitem[{Wade et~al.(2018)Wade, Coniglio,, and Ziegler}]{wade2018comparison}
Wade, A.~R., M.~C. Coniglio, and C.~L. Ziegler, 2018: Comparison of near-and far-field supercell inflow environments using radiosonde observations. \textit{Mon.\ Wea.\ Rev.}, \textbf{146~(8)}, 2403--2415.

\bibitem[{Wade et~al.(2023)Wade, Jirak,, and Lyza}]{wade2023regional}
Wade, A.~R., I.~L. Jirak, and A.~W. Lyza, 2023: Regional and seasonal biases in convection-allowing model forecasts of near-surface temperature and moisture. \textit{Wea.\ Forecasting}, \textbf{38~(12)}, 2415--2426.

\bibitem[{Warren et~al.(2017)Warren, Richter, Ramsay, Siems,, and Manton}]{warren2017impact}
Warren, R.~A., H.~Richter, H.~A. Ramsay, S.~T. Siems, and M.~J. Manton, 2017: Impact of variations in upper-level shear on simulated supercells. \textit{Mon.\ Wea.\ Rev.}, \textbf{145~(7)}, 2659--2681.

\bibitem[{Warren et~al.(2021)Warren, Richter,, and Thompson}]{warren2021spectrum}
Warren, R.~A., H.~Richter, and R.~L. Thompson, 2021: Spectrum of near-storm environments for significant severe right-moving supercells in the contiguous {U}nited {S}tates. \textit{Mon.\ Wea.\ Rev.}, \textbf{149~(10)}, 3299--3323.

\bibitem[{Werkema(2022)}]{werkema2022multistorm}
Werkema, A.~D., 2022: Mutual influences of adjacent supercells in multistorm simulations. Ph.D. thesis, North Carolina State University.

\bibitem[{Wilks(2011)}]{wilks2011statistical}
Wilks, D.~S., 2011: \textit{Statistical Methods in the Atmospheric Sciences}, Vol. 100. Academic press.

\bibitem[{Wilson et~al.(2023)}]{wilson2023environmental}
Wilson, M.~B., and Coauthors, 2023: Environmental and storm-scale controls on close proximity supercells observed by {TORUS} on 8 {J}une 2019. \textit{Mon.\ Wea.\ Rev.}, \textbf{151~(12)}, 3013--3035.

\end{thebibliography}

\end{document}